\documentclass[twocolumn,prd,superscriptaddress,preprintnumbers,amsmath,amssymb,nofootinbib]{revtex4-1}
\usepackage{amssymb}
\usepackage{slashed}
\usepackage{graphicx}
\usepackage{graphics}
\usepackage{epsfig}
\usepackage{color}
\usepackage{soul}
\usepackage{tabularx}
\usepackage{amsmath}
\usepackage{amsfonts}
\usepackage{amssymb}
\usepackage{hyperref}
\usepackage{microtype}

\newcommand{\be}{\begin{equation}}
\newcommand{\ee}{\end{equation}}
\newcommand{\bea}{\begin{eqnarray}}
\newcommand{\eea}{\end{eqnarray}}

\begin{document}
\author{Nicolai Christiansen}
\email[]{n.christiansen@thphys.uni-heidelberg.de}
\affiliation{Institut f\"ur Theoretische
  Physik, Universit\"at Heidelberg, Philosophenweg 16, 69120
  Heidelberg, Germany}

\author{Astrid Eichhorn}
   \email[]{a.eichhorn@thphys.uni-heidelberg.de}
\affiliation{Institut f\"ur Theoretische
  Physik, Universit\"at Heidelberg, Philosophenweg 16, 69120
  Heidelberg, Germany}

\title{An asymptotically safe solution to the U(1) triviality problem}

\begin{abstract}
We explore whether quantum gravity effects within the asymptotic safety paradigm can provide a predictive ultraviolet completion for Abelian gauge theories. We evaluate the effect of quantum gravity fluctuations on the running couplings in the gauge sector and discover an asymptotically safe fixed point of the Renormalization Group. In particular,  if the strength of gravitational interactions remains below a critical strength, the minimal gauge coupling becomes asymptotically free.
Further, we point out that  a completely asymptotically free 
dynamics for the gauge field is impossible to achieve, as asymptotically safe 
quantum gravity necessarily induces nonvanishing higher-order interactions for 
the gauge field in the ultraviolet.
\end{abstract}

\maketitle

\section{The asymptotic safety paradigm for gauge-gravity systems}
Here, we explore the scenario that the coupling of a U(1) gauge theory to 
asymptotically safe quantum gravity \cite{Weinberg:1980gg} induces a predictive 
ultraviolet completion for the gauge theory. In particular, 
this could provide a solution to the triviality problem in the Abelian sector of 
the Standard Model 
\cite{GellMann:1954fq,Gockeler:1997dn,Gockeler:1997kt,Gies:2004hy}, and would 
thus constitute an intriguing alternative to models of grand unification. 
Specifically, we find evidence suggesting that under the impact of quantum 
gravitational fluctuations, Abelian gauge theories develop an ultraviolet (UV) 
fixed point, as first discussed in \cite{Harst:2011zx}. In principle, one 
scenario for such a fixed point could be that the gauge theory becomes 
completely asymptotically free, and all gauge interactions vanish in the 
ultraviolet. Here, we will show that the UV completion for gauge theories 
induced by asymptotically safe quantum gravity is a little more intricate. 
In particular,
higher-order interactions play an important role: Higher-order field-strength invariants such as $\left(F_{\mu \nu}F^{\mu 
\nu}\right)^2$ are necessarily present in the ultraviolet, 
i.e., quantum gravity induces interactions for Abelian gauge fields even in the 
absence of charged fields. This has a crucial impact on the Renormalization 
Group (RG) flow: While the beta-function for a pure U(1) gauge 
theory without 
 charged matter
is of course zero in the absence of gravity, the 
presence of gravity-induced higher-order operators means that the gauge 
couplings run as a function of the momentum scale. Moreover, an interacting RG fixed point for gravity necessarily requires the 
coupling of the higher-order operator $\left(F_{\mu \nu}F^{\mu \nu}\right)^2$ to 
be nonzero. 
These couplings contribute to rendering the U(1) gauge coupling $e$ either 
asymptotically free or safe in the ultraviolet. In both cases, complete 
asymptotic freedom for the gauge sector is impossible to achieve, as even in the 
case of an asymptotically free gauge coupling $e$, the couplings of 
higher-order gauge interactions must remain nonzero, and thus must become 
asymptotically safe. 

As has been explored for the case of scalars and fermions in the context of 
asymptotic safety before, quantum gravity necessarily induces new interactions 
\cite{Eichhorn:2011pc,Eichhorn:2011ec,Eichhorn:2012va,Eichhorn:2013ug,
Meibohm:2016mkp,Eichhorn:2016esv,Eichhorn:2017eht}. 
These interactions are those 
that are compatible with the symmetries of the kinetic terms of the respective 
fields, see \cite{Eichhorn:2017eht,Eichhorn:2016esv}. For the case of gauge 
fields, the 
whole tower of gauge-invariant interactions can be induced by quantum gravity. 
For the minimal gauge coupling $e$ with vanishing canonical 
dimension, the Ward-Takahashi identities imply a relation between the 
$\beta$-function of the coupling and the anomalous dimension  $\eta_A$ of the  
gauge field
\be
\beta_{e^2} =  e^2 \, \eta_A.
\ee
 Quantum-gravity corrections enter $\eta_A$ and imply that this 
coupling features a fixed point at zero in the presence of gravity. 
In the presence of charged matter, the free fixed point is ultraviolet repulsive, leading to the triviality problem. Quantum gravity can alter that scaling behavior, and thus provide a solution.

In this paper, we have three main goals
\begin{itemize}
\item We will show that higher-order gauge interactions are non-vanishing in a 
quantum-gravity induced ultraviolet completion for an Abelian gauge theory.
\item We will point out that the requirement that the fixed point lies at real values of the couplings imposes bounds on the viable gravitational parameter space.
\item We will study the impact of induced interactions on the canonical gauge coupling to determine whether it is rendered relevant or irrelevant in the ultraviolet. In the latter case, the Standard Model value for the U(1) hypercharge coupling at the Planck scale would be difficult to reconcile with an asymptotically safe UV completion.
\end{itemize}

\section{Functional Renormalization group for gauge-gravity systems}
We will employ the functional Renormalization Group which allows us to probe the 
scale dependence of a quantum field theory of gravity and matter. Specifically, 
the Wetterich equation encodes the response of the theory to 
the change of a momentum scale $k$: As $k$ is changed, quantum fluctuations 
with momenta $p^2 \approx k^2$ yield the main contribution to the change of 
the effective dynamics, which is encoded in scale dependent running couplings. 
Technically, this coarse-graining procedure is implemented with the help of a 
cutoff function $R_k(p^2)$. Specifically, we will choose a Litim 
cutoff \cite{Litim:2001up}
\be
R_k = Z_k\left(k^2 -p^2\right)\theta(k^2-p^2).
\ee 
The full, regularized and scale-dependent propagator of quantum fluctuations is given by
$\left(\Gamma_k^{(2)}+R_k \right)^{-1}$,
where $\Gamma_k^{(2)}$ denotes the second functional derivative of the scale-dependent effective action with respect to the fields and $R_k$ is the shape function for the infrared cutoff term. With these ingredients, the Wetterich equation \cite{Wetterich:1992yh}, see also \cite{Morris:1993qb,Ellwanger:1993mw} is given by
\be
\partial_t \Gamma_k = \frac{1}{2}{\rm Tr}\left(\Gamma_k^{(2)}+R_k \right)^{-1}\partial_t R_k,
\ee
where $\partial_t = k \, \partial_k$. For reviews see \cite{Berges:2000ew, Polonyi:2001se,
Pawlowski:2005xe, Gies:2006wv, Delamotte:2007pf, Rosten:2010vm, Braun:2011pp}. For the gravity fluctuations, we will employ a linear split
\be
g_{\mu \nu}= \bar{g}_{\mu \nu}+ h_{\mu \nu},
\ee
which allows us to write a background-field gauge fixing as well as a regulator 
term that depend on the background metric $\bar{g}_{\mu \nu}$. We thereby arrive 
at a regularized propagator for the fluctuation field $h_{\mu \nu}$. The beta 
functions for the gauge field are independent of the choice of background metric 
$\bar{g}_{\mu\nu}$, and we employ the minimal choice $\bar{g}_{\mu \nu} = 
\delta_{\mu \nu}$. Following Reuter's groundbreaking work \cite{Reuter:1996cp}, 
substantial evidence for the viability of the asymptotic safety paradigm in 
quantum gravity has been collected in  
\cite{Reuter:2001ag,Lauscher:2001ya,Lauscher:2002sq,Litim:2003vp,Fischer:2006fz,
Codello:2006in,Machado:2007ea,Codello:2008vh,Eichhorn:2009ah,Manrique:2009uh,
Benedetti:2009rx,Eichhorn:2010tb,Groh:2010ta,Manrique:2010mq,Manrique:2010am,
Manrique:2011jc,Benedetti:2012dx,Christiansen:2012rx,Dietz:2012ic,Falls:2013bv,
Christiansen:2014raa,Becker:2014qya,Falls:2014tra,Eichhorn:2015bna,
Demmel:2015oqa,Christiansen:2015rva,Gies:2015tca,Morris:2016spn,Percacci:2016arh,Gies:2016con,Ohta:2016npm,Henz:2016aoh,Biemans:2016rvp,Pagani:2016dof,Christiansen:2016sjn,Denz:2016qks,Ohta:2017dsq,Falls:2017cze}, also in the case with matter  
\cite{Benedetti:2009gn,Harst:2011zx,Eichhorn:2011pc,Eichhorn:2011ec,
Eichhorn:2012va,Dona:2012am,Eichhorn:2013ug,Dona:2013qba,Dona:2014pla,
Oda:2015sma,Meibohm:2015twa,Dona:2015tnf,Meibohm:2016mkp,Eichhorn:2016esv,
Eichhorn:2017eht,Biemans:2017zca}, see, e.g., \cite{ASreviews} for reviews. 
Consequences in astrophysics and cosmology have been explored, e.g., in 
\cite{cosmology}.

As quantum fluctuations generate all field monomials compatible with the symmetries, the Wetterich equation provides a flow in the infinite dimensional space of all couplings. For practical calculations, a truncation of the effective dynamics is necessary.
Here, we follow the assumption that canonical scaling provides a good guideline 
for the setup of truncations in asymptotically safe gravity 
with matter. This assumption is justified if the interacting fixed point 
underlying asymptotic safety is not strongly non-perturbative, as then canonical 
scaling determines the main contribution to the critical exponents. Within the 
asymptotic safety paradigm for gravity as well as matter, the assumption appears 
to be justified 
\cite{Falls:2013bv,Falls:2014tra,Narain:2009fy,Narain:2009gb,Eichhorn:2016esv, 
Eichhorn:2016vvy}.

\begin{figure}[t!]
\includegraphics[width=0.3\linewidth]{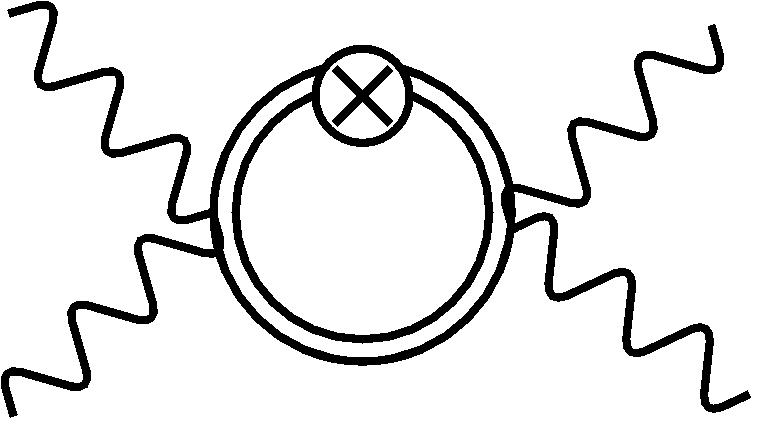}\quad \includegraphics[width=0.28\linewidth]{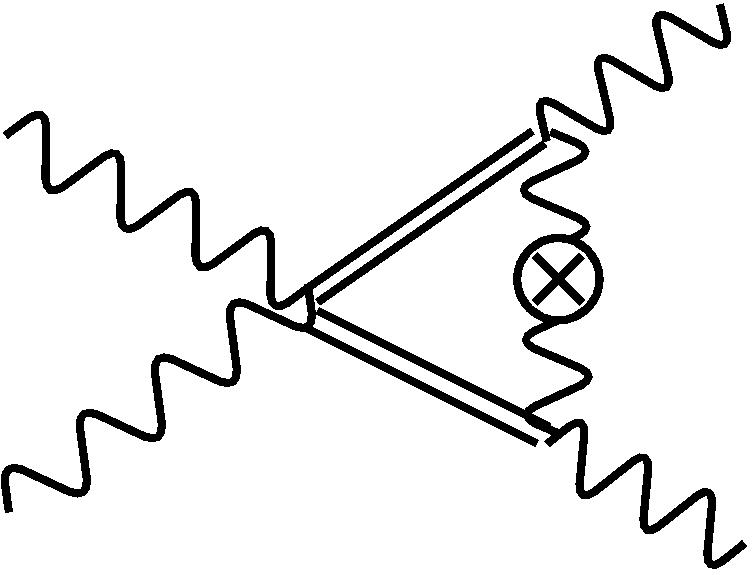}\quad \includegraphics[width=0.32\linewidth]{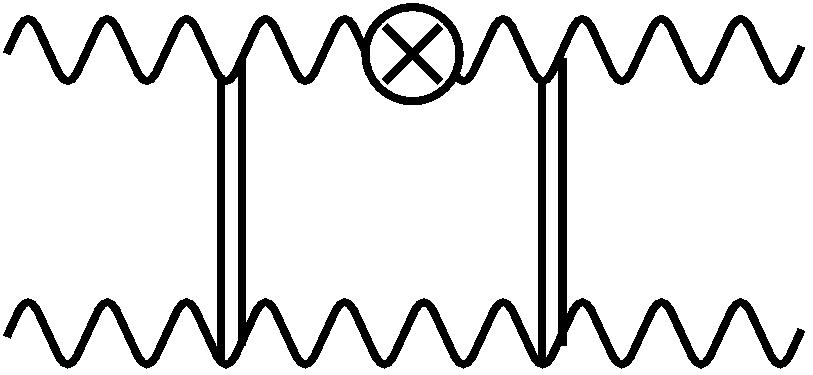}\newline\\
\includegraphics[width=0.2\linewidth]{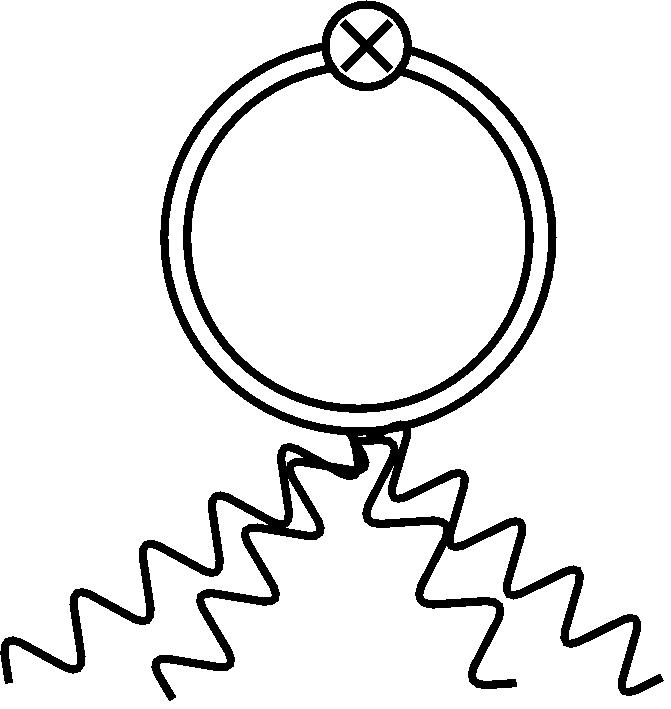}\quad\includegraphics[width=0.32\linewidth]{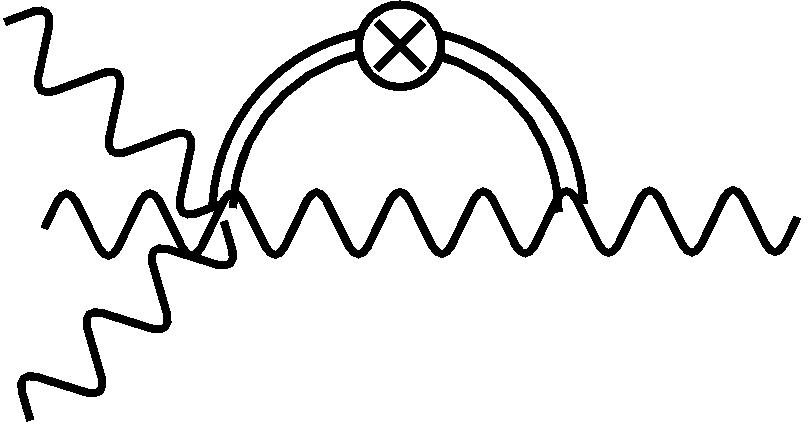}\quad \includegraphics[width=0.28\linewidth]{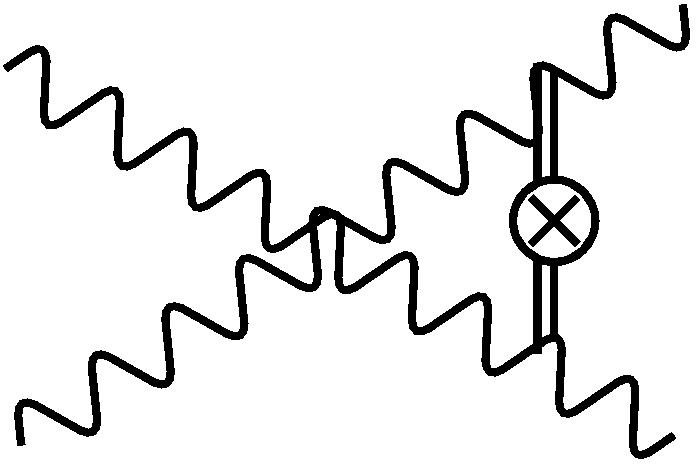}\quad \newline\\
 \includegraphics[width=0.3\linewidth]{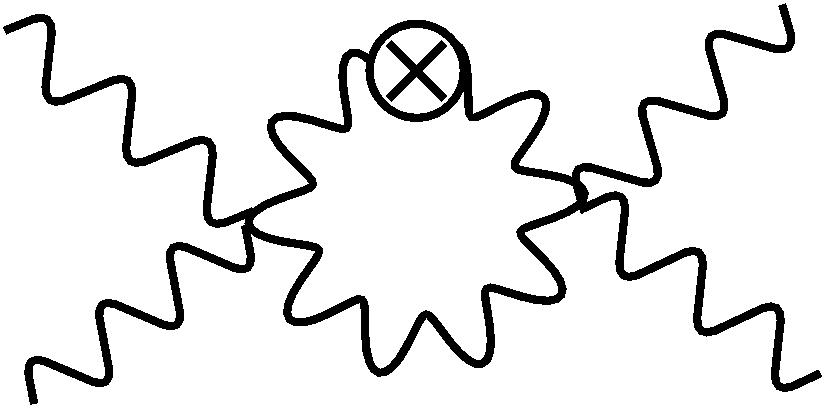}
\caption{\label{fig:w2diags} These diagrams contribute to the flow of $w_2$. Curly lines denote photons and double lines metric fluctuations. The regulator insertion $\partial_t R_k$ is denoted by a crossed circle, and can in turn be found on any of the internal lines of a diagram. We only show one representative of each class. The first line contains all diagrams $\sim g^2$, the second line diagrams $\sim g\, w_2$ and the third line the diagram $\sim w_2^2$.}
\end{figure}

\begin{figure}[t!]
\includegraphics[width=0.15\linewidth]{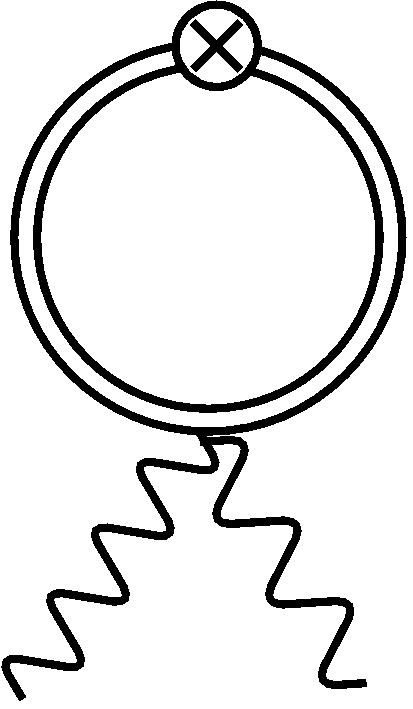}\quad\quad \includegraphics[width=0.3\linewidth]{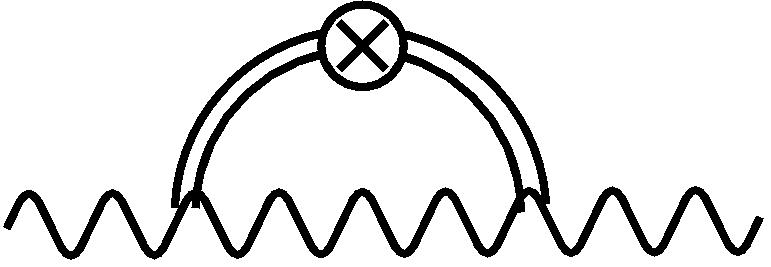}\quad \quad\includegraphics[width=0.15\linewidth]{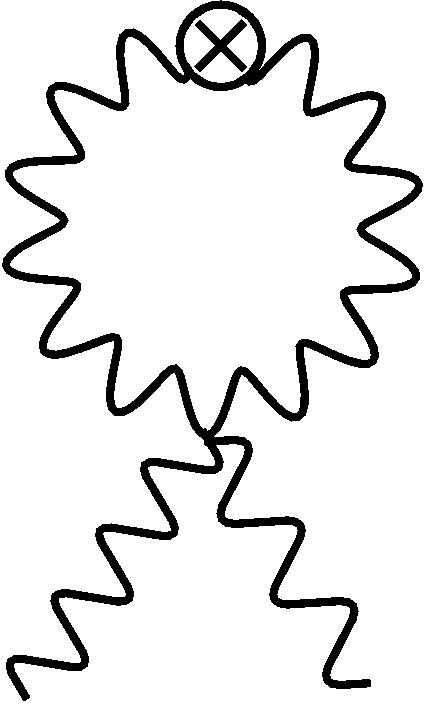}
\caption{\label{fig:etadiags} These diagrams contribute to the flow of $\eta_A$. The diagram to the left and in the middle are the gravitational contribution to $\eta_A$, whereas the diagram to the right is the pure-gauge contribution $\sim w_2$.}
\end{figure}

We will focus on a minimal truncation in the gauge sector that exhibits how 
asymptotically safe quantum gravity induces an interacting fixed point in the 
gauge sector. We discuss how this might provide a mechanism  to solve
the triviality problem. 
To that end, it is sufficient to include the 
first higher-order operator in the gauge sector. A more 
extended truncation will be studied elsewhere.
Accordingly, our truncation is
\bea
\Gamma_k &=& \frac{Z_A}{4}\int d^4x\, \sqrt{g}\, g^{\mu \nu}g^{\kappa 
\lambda}F_{\mu \kappa}F_{\nu \lambda} \nonumber\\
&{}&+ \frac{\bar{w}_2}{8} \int d^4x\, \sqrt{g}\, \left(g^{\mu \nu}g^{\kappa 
\lambda}F_{\mu \kappa}F_{\nu \lambda} \right)^2 \nonumber\\
&{}& + \Gamma_{k\, \rm EH} + S_{{\rm gf},\, h}+ S_{{\rm gf},\, A}.
\eea
Note that all couplings are $k$-dependent, which we do not indicate explicitly for notational simplicity. In the above, the Einstein-Hilbert truncation for the metric fluctuations is obtained from
\be
\Gamma_{k\, \rm EH}= \frac{-1}{16 \pi \bar{G}_N}\int d^4x\, \sqrt{g}\, \left( R- 2 \bar{\Lambda}\right),
\ee
by expanding to second order in $h_{\mu \nu}$ and then redefining the field according to
\be
h_{\mu \nu} \rightarrow \sqrt{Z_h\, \bar{G}_N}h_{\mu \nu}.
\ee
The two-gauge fixing terms are given by
\bea
S_{{\rm gf},\, h}&=& \frac{Z_h}{\alpha_h\, 32 \pi}\int d^4x\sqrt{\bar{g}}\,\bar{g}^{\mu \nu} \mathcal{F}_{\mu}\mathcal{F}_{\nu},\\
\mathcal{F}_{\mu} &=& \bar{D}^{\nu}h_{\mu \nu} - \frac{1+\beta}{4}\bar{D}_{\mu}h,\\
S_{{\rm gf},\, A}&=& \frac{Z_A}{2 \alpha_A}\int d^4x\, 
\sqrt{\bar{g}}\left(\bar{D}^{\mu}A_{\mu} \right)^2.
\eea
In the following, we will focus on Landau gauge for the photons, where $\alpha_A =0$, and on $\beta=1, \, \alpha=0$ for the metric.
We introduce dimensionless couplings as follows:
\bea
w_2 &=&\frac{\bar{w}_2}{Z_A^2} k^4,\quad
g=\frac{\bar{G}_N}{Z_h} k^2,\, \quad \mu_h = -2\bar{\Lambda} Z_h\, k^{-2}.
\eea
Moreover the anomalous dimensions are given by
\be
\eta_h= -\partial_t \ln Z_h, \quad \eta_A = - \partial_t \ln Z_A.
\ee
Diagrammatically, the flow of the coupling $w_2$ is encoded in the diagrams in Fig.~\ref{fig:w2diags}, and that of the anomalous dimension in the diagrams in Fig.~\ref{fig:etadiags}.
\begin{widetext}
To project the RG flow obtained with the help of the Wetterich equation onto $\bar{w}_2$, we employ the following rule:
\bea
\partial_t \bar{w}_2 =\underset{p^2\rightarrow0}{\rm lim} 
\frac{1}{720\, (p^2)^2} P_{\mu 
\nu\kappa \lambda}(p,p,-p,-p) \left(\frac{\delta}{\delta A_{\mu}(p_1)} 
\frac{\delta}{\delta A_{\nu}(p_2)} \frac{\delta}{\delta A_{\kappa}(p_3)} 
\frac{\delta}{\delta A_{\lambda}(p_4)} \partial_t 
\Gamma_k\right)\Big|_{p_1=p_2=-p_3=-p_4=p, A=0,h=0 }
\eea
\end{widetext}
Herein, the tensor $P_{\mu \nu\kappa \lambda}(p)$ is defined as
\bea
&{}&P_{\mu \nu\kappa \lambda}(p_1,p_2,p_3,p_4)\\
&{}&= \left(\frac{\delta}{\delta A_{\mu}(p_1)} \frac{\delta}{\delta A_{\nu}(p_2)} \frac{\delta}{\delta A_{\kappa}(p_3)} \frac{\delta}{\delta A_{\lambda}(p_4)} \frac{1}{8}\left(F^2\right)^2\right).\nonumber
\eea

Similarly, we project onto the anomalous dimension for the photon as follows
\be
\eta_A = -\underset{p^2\rightarrow0}{\rm lim} \frac{1}{Z_A p^2} P_{\mu\nu}(p)\left(\frac{\delta}{\delta A_{\mu}(p)}\frac{\delta}{\delta A_{\nu}(-p)} \partial_t\Gamma_k\right)\Big|_{ A=0,h=0 },
\ee
where 
\be
P_{\mu\nu} (p)= \frac{1}{3}\left(\delta_{\mu \nu}- \frac{p_{\mu}p_{\nu}}{p^2}\right).
\ee
We will not evaluate the running of the gravitational couplings, instead treating them as free parameters. Thereby we test how the system will respond to extensions of the truncation in the gravitational sector, which could shift the fixed-point values of the gravitational couplings.

\section{Results: Quantum-gravity induced asymptotic safety in the gauge sector}
\subsection{Shifting the Gau\ss{}ian fixed point}
At vanishing gravitational coupling $g=0$, the system exhibits 
a non-interacting, i.e., Gau\ss{}ian fixed point, at which $w_2=0$. Moreover,  
$w_2$ corresponds to an UV-irrelevant direction, according to its canonical 
dimension $-4$.
 In the presence of quantum gravity, there are several 
diagrams that shift the Gau\ss{}ian fixed point to an interacting one. In 
other words, these diagrams are independent of the coupling $w_2$, and are nonzero as soon as the canonical field-strength-squared interaction for photons is minimally coupled to gravity. Thus, higher-order gauge interactions are automatically induced in the ultraviolet.
Depending on the sign and magnitude of the gravitational contribution, the fixed point might even get shifted into the complex plane. In that case, asymptotically safe gravity cannot be reconciled with the existence of a  fundamental Abelian gauge sector in the Standard Model. 

Specifically, these are the diagrams in the first line of Fig.~\ref{fig:w2diags}. Together, we call this contribution to $\beta_{w_2}$, which is independent of $w_2$ itself, but proportional to $g^2$, the induced contribution, as it induces an interacting fixed point.
It contains the following contributions in the order in which they are shown in Fig.~\ref{fig:w2diags}.

\bea
\beta_{w_2}\Big|_{\rm induced}&=& \frac{8}{3 \left(1+\mu_h\right)^3}g^2(6-\eta_h)\label{eq:betaw2induced}\\
&-&g^2\left(2\frac{8-\eta_h}{(1+\mu_h)^3} +\frac{8-\eta_A}{(1+\mu_h)^2}\right)\nonumber\\
&+&g^2 4\left(\frac{10-\eta_h}{5(1+\mu_h)^3} +\frac{10-\eta_A}{5(1+\mu_h)^2}\right).\nonumber
\eea
These contributions are responsible for the main effect that we highlight in this work: All terms in eq.~\eqref{eq:betaw2induced} are nonzero even if we set $w_2=0$. Therefore this value of the coupling is \emph{not} a fixed point, i.e., its beta-function is nonzero at this point. Accordingly the only possible fixed point for the gauge sector in the presence of quantum gravity, i.e., for $g\neq 0$, is an interacting one. Thus, as $g$ is increased starting from zero, the Gau\ss{}ian fixed point at $w_2=0$ is shifted to become an interacting one.

\subsection{Asymptotic safety for U(1) gauge theories}

Further, there are contributions to the beta-function which are $\sim w_2 \, g$. We call them mixed, as they only exist for nonzero $g$ and $w_2$.
These can be understood as a quantum-gravity correction to the scaling dimension of the coupling $w_2$. In the order in which they appear in the second line in Fig.~\ref{fig:w2diags}, they are given by
\bea
\beta_{w_2}\Big|_{\rm mixed}&=& - \frac{2}{3 \pi \left( 1+\mu_h\right)^2}g\, w_2\, (6- \eta_h)\\
&+&g\, w_2 \left(\frac{8-\eta_h}{2\pi(1+\mu_h)^2}+\frac{8-\eta_A}{2\pi(1+\mu_h)} \right)\nonumber\\
& -&\frac{13 g\, w_2}{60\pi}\left(\frac{10-\eta_h}{(1+\mu_h)^2}+2\frac{10-\eta_A}{1+\mu_h} \right).\nonumber
\eea

The pure-photon-diagram in the 3rd line of Fig.~\ref{fig:w2diags} yields
\be
\beta_{w_2}\Big|_{\rm pure-photon} = \frac{19}{240 \pi^2}(10 -\eta_A)w_2^2.
\ee
If the theory is  coupled to charged matter 
fields, there will be contributions of box diagrams, which are all of 
order $e^4$. In particular, these will not contribute to the fixed-point value 
nor the critical exponents at the shifted Gaussian fixed point characterized 
by $e_*=0$.
The complete beta function for $w_2$ is given by
\bea
\nonumber \beta_{w_2}&=& 4 w_2 + 2 \eta_A\, w_2 + e^4 \beta_{w_2}\Big|_{\rm matter} 
\label{eq:betaw2full} \\
&+& \beta_{w_2}\Big|_{\rm induced} + \beta_{w_2}\Big|_{\rm mixed}+ \beta_{w_2}\Big|_{\rm pure-photon}.
\eea

For the gauge coupling $e$, we obtain
\bea
\beta_{e^2}& =& e^2 \,\eta_A \nonumber\\
&=& e^2 \Bigl( \frac{1}{4\pi(1+\mu_h)^2}g(6-\eta_h)\nonumber\\
&{}&\quad\quad- \frac{g}{8\pi} \left(\frac{8-\eta_h}{(1+\mu_h)^2} + \frac{8-\eta_A}{1+\mu_h}\right)\nonumber\\
&{}& \quad\quad- \frac{w_2}{24\pi^2}(8-\eta_A) \, + 
\Delta \eta_{A,\mathrm{matter}} e^2  \Bigr), \label{eq:etaA}
\eea
where we have parameterized the contributions of charged 
matter as $\Delta \eta_{A,\mathrm{matter}} e^4$, with $\Delta \eta_{A, \rm matter}>0$.

\begin{figure}[t!]
\includegraphics[width=\linewidth]{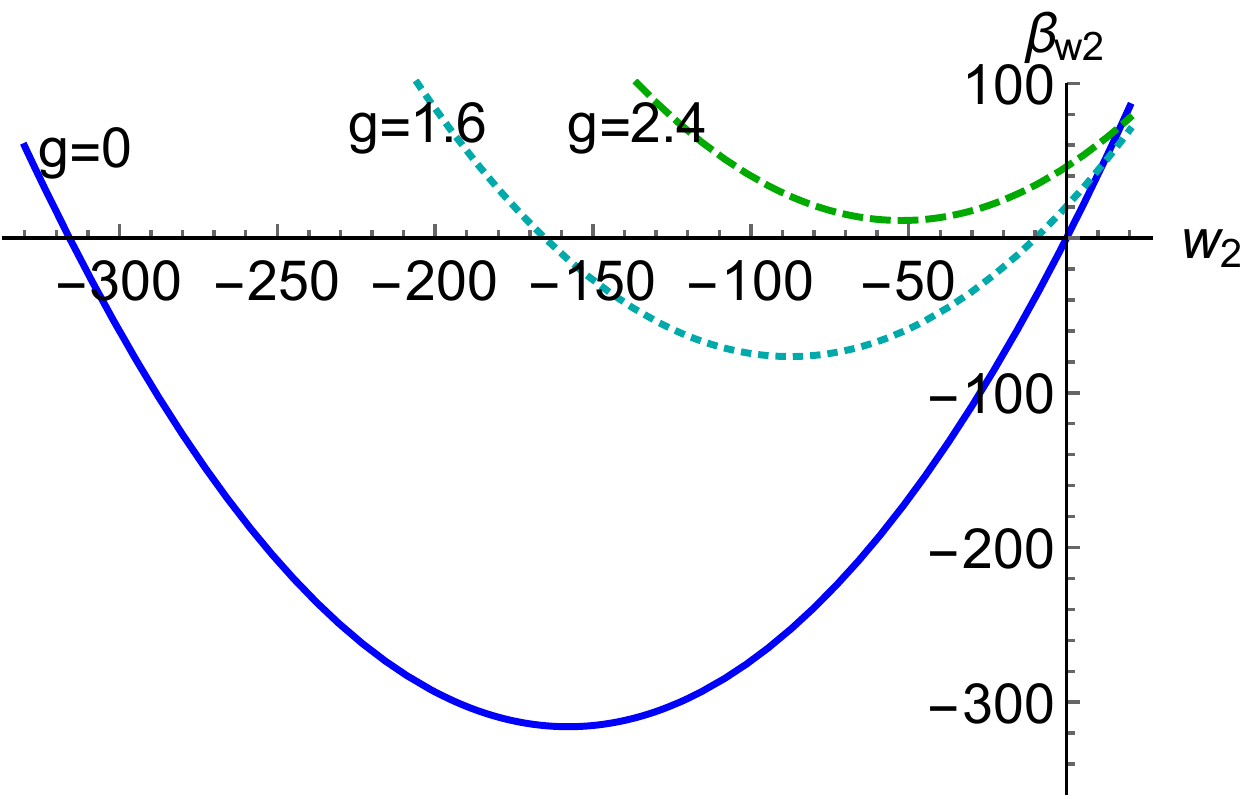}
\caption{\label{fig:beta_w2_g} We show $\beta_{w_2}$ for $\eta_h=0=\eta_A$ as well as $\mu_h=0$ for different values of $g$. Any finite value $g>0$ shifts the Gau\ss{}ian to an interacting fixed point.
Beyond a critical value of $g$, no real fixed point exists.}
\end{figure}

From \eqref{eq:etaA} we infer that 
the gravity contributions dominate 
for small gauge 
coupling $e$, as they couple 
linearly in $e^2$. 
In comparison, the contributions from matter are of order $e^4$ 
and therefore do not contribute to the scaling exponent at the fixed point at 
$e_{\ast}=0$. In the absence of gravity, these contributions induce a triviality problem. Interestingly, they become subdominant in the presence of gravity, and can thus be neglected when exploring whether quantum gravity renders Abelian gauge theories asymptotically free in the gauge coupling $e$. The main quantum-gravity effects on the Abelian gauge coupling can be understood within a simplified setting, where we set the anomalous dimensions to zero in the loop integrals and also neglect the cosmological constant. In that case, we have
\bea
\beta_{e^2}|_{{\rm loops:}\, \eta=0}&=& \left(-\frac{g}{2\pi}- \frac{w_2}{3 
\pi^2}\right)e^2, \label{eq:betaalpha_simp}\\
\beta_{w_2}|_{{\rm loops:}\, \eta=0}&=& 4 w_2 + 8 g^2 - \frac{7}{2\pi}g\, w_2 + \frac{1}{8\pi^2}w_2^2.\nonumber\\
&{}& \label{eq:betaw2_simp}
\eea
 From eq.~\eqref{eq:betaalpha_simp}, it is clear that quantum gravity effects 
can render the U(1) gauge coupling asymptotically free, providing a solution to 
the triviality problem, as discussed in \cite{Harst:2011zx}. Here, we will 
highlight that this scenario hinges on the weakness of the gravitational 
coupling, due to an intriguing interplay of the direct contribution ($\sim g$) 
and the mediated one ($\sim w_2$) in eq.~\eqref{eq:betaalpha_simp}.  Note that 
the addition of charged matter does not change the mechanism for gravity-induced 
asymptotic freedom, as charged matter contributes to the beta function for the 
gauge coupling at higher order in $e$.

Without gravity, $w_2$ features a non-interacting fixed point, at which it is irrelevant according to canonical power counting. With gravity, the situation changes: The term $\sim 8 g^2$ prevents the possibility of a free fixed point for the coupling $w_2$. Instead, the fixed point is shifted towards increasingly negative values as a function of $g$, 
\bea
w_{2\ast\, \rm sGFP}\!=\! 2\pi\! \left(\!7g - 8 \pi\! +\! \sqrt{64\pi^2-112g\, \pi +33g^2} \right).\label{eq:w2sGFP}
\eea
Note that our expansion of the full ``potential" of the gauge field, $V(F^2)$ to 
second order is insufficient to determine global stability properties. Thus, a 
non-perturbative phenomenon like the formation of  a nontrivial minimum in 
$V(F^2)$ in the UV, as well as a destabilization in the sense of a microscopic 
potential that is not bounded from below, are both in principle possible, and 
might seem indicated by the negative fixed-point value for $w_2$. To settle this 
question, studies of $V(F^2)$ are necessary. These are feasible along the lines 
of, e.g., \cite{Gies:2002af,Eichhorn:2010zc}.

Since the fixed point in Eq.~\eqref{eq:w2sGFP} emerges from the non-interacting, Gau\ss{}ian fixed point, as we increase $g$, we call it the shifted Gau\ss{}ian fixed point. In particular, its UV-attractivity properties remain similar to that of the Gau\ss{}ian fixed point, i.e., $w_2$ remains irrelevant at that fixed point. Accordingly the critical exponent remains negative
\be
\theta_{w_2}= - \frac{\partial \beta_{w_2}}{\partial w_2}\Big|_{w_2=w_{2 \, \rm sGFP}}= - \frac{1}{2\pi}\sqrt{64\pi^2 -112\pi\, g + 33g^2}. 
\ee
It is obvious that once $g$ exceeds a critical value $g_{\rm crit}$, the critical exponent vanishes and $w_{2\ast\, \rm sGFP}$ becomes complex. At this point, the shifted Gau\ss{}ian fixed point annihilates with the second zero of $\beta_{w_2}$ and the two move off into the complex plane, cf.~Fig.~\ref{fig:beta_w2_g}. For rather large values $g\approx 8.4$, real fixed points reappear. 
For these, the back-coupling of $w_2$ into $\beta_{e^2}$ 
induces a huge departure from canonical scaling, invalidating the rationale 
behind our truncation. 
 Moreover,
fixed-point values for gravity typically do not lie in that regime, and 
we thus focus on the significantly more important bound on $g$ that arises from 
the shift of the fixed-point values into the complex plane.

\begin{figure}[t!]
\includegraphics[width=\linewidth]{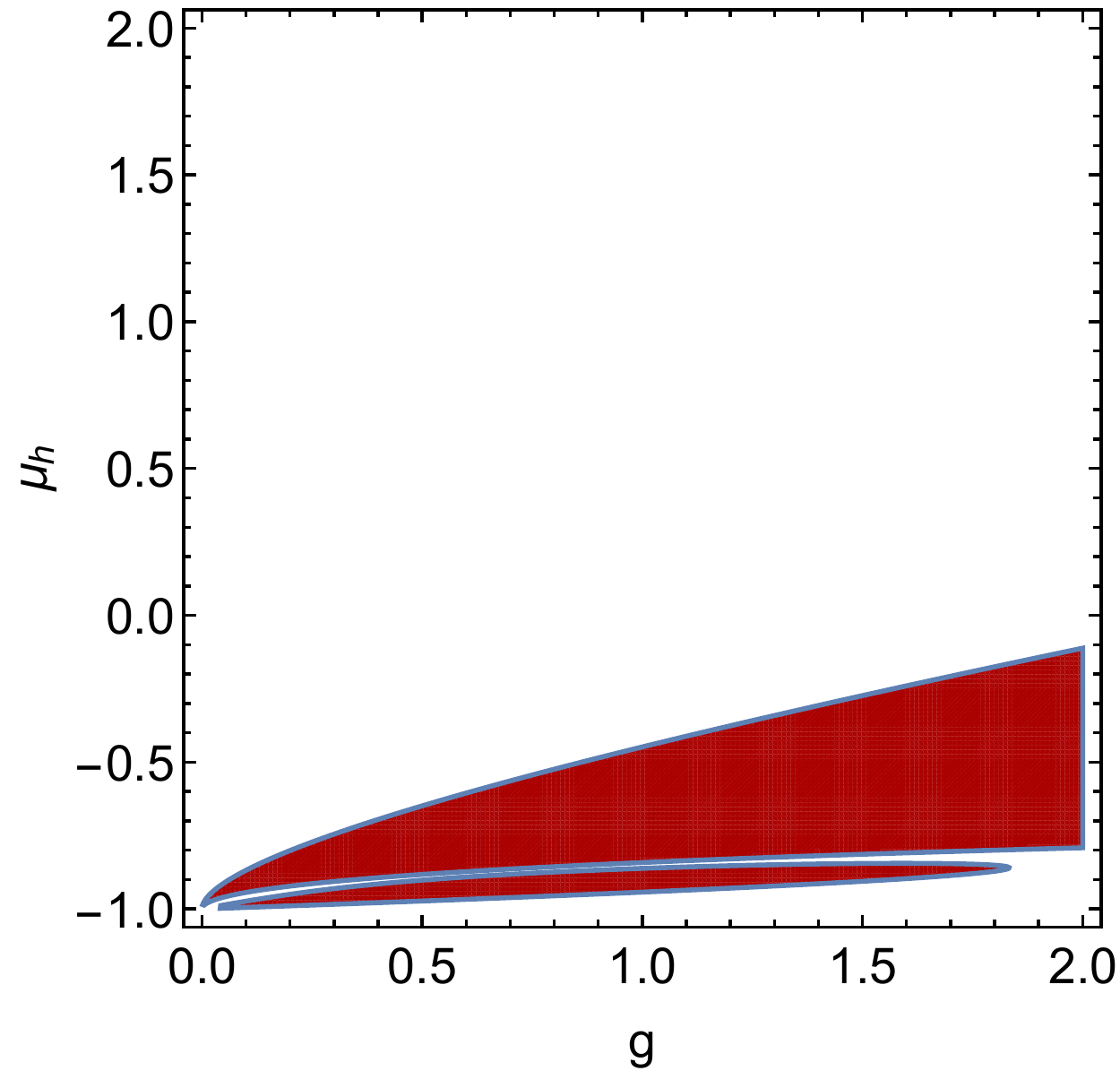}
\caption{\label{fig:exclusions} No real fixed point exists in the red region in the gravitational parameter space for $\eta_h=0$, i.e., the white region is allowed. The second allowed region at negative $\mu_h$ is a new fixed point that appears to be induced by a pole in the beta function.}
\end{figure}

Analyzing the full beta functions eq.~\eqref{eq:betaw2full} instead of the simplification in eq.~\eqref{eq:betaw2_simp}, requires us to use the full anomalous dimension,  see eq.~\eqref{eq:etaA}
\be
\eta_A = -\frac{8 w_2 (1+\mu_h)^2 + 3\pi\, g\, (4+8 \mu_h + 
\eta_h)}{(1+\mu_h)((1+\mu_h)(24 \pi^2 -w_2)- 3\pi\, g)}.\label{eq:etaAfull}
\ee
Confirming our analysis of the simplified beta functions 
Eq.~\eqref{eq:betaw2_simp} and Eq.~\eqref{eq:betaalpha_simp} no real fixed point 
exists for a certain region of the gravitational parameter space, 
cf.~Fig.~\ref{fig:exclusions}; excluding this region from the viable fixed-point 
parameter space.  The bounds in Fig.~\ref{fig:exclusions} only show a very mild dependence on $\eta_h$, e.g., for $g=1$ and $|\eta_h|=0.5$ there is only a $5 \% $ deviation in the 
critical value of $\mu_h$. In general, for values of $\eta_h >0$ the red region in Fig. 
\ref{fig:exclusions} shrinks, while the opposite is true for $\eta_h <0$.

As a next step, we evaluate the quantum-gravity effects on the gauge coupling 
$e^2$. There is firstly a direct effect, i.e., loop diagrams with internal 
metric fluctuations induce a nontrivial running of the coupling.  These 
contribute to render the gauge coupling asymptotically free, thus providing a 
solution to the triviality problem.
Secondly, the quantum-gravity induced coupling $w_2$ also contributes to the 
running of $e^2$. The second effect is an \emph{indirect} quantum-gravity 
effect, i.e., it is mediated by gauge interactions. These interactions are 
induced by quantum-gravity fluctuations and then in turn affect the scaling 
dimension of the gauge coupling. Thus we call these the mediated contribution.
The contributions to the critical exponent of the gauge 
coupling can be split accordingly,
\begin{equation}
\theta_{e^2} = - \frac{\partial \beta_{e^2}}{\partial 
e^2}\Big|_{e=0} 
=\theta_{e^2,\,\rm direct}+\theta_{e^2,\,\rm mediated},
\end{equation}
In the approximation given by \eqref{eq:betaalpha_simp} and 
\eqref{eq:betaw2_simp} one obtains
\bea
&{}& \theta_{e^2,\,\rm direct}=\theta_{e^2,\, g}= \frac{g}{2\pi}>0, 
\label{eq:thetaalphag}\\
&{}&\theta_{e^2,\,\rm mediated}= \theta_{e^2, \, 
w_2}=\frac{w_{2\, \rm sGFP}}{3 \pi^2}<0.\label{eq:thetaalphaw2}
\eea
Intriguingly, the two contributions have the opposite sign, which has 
important consequences.
\begin{figure}[!t]
\includegraphics[width=\linewidth]{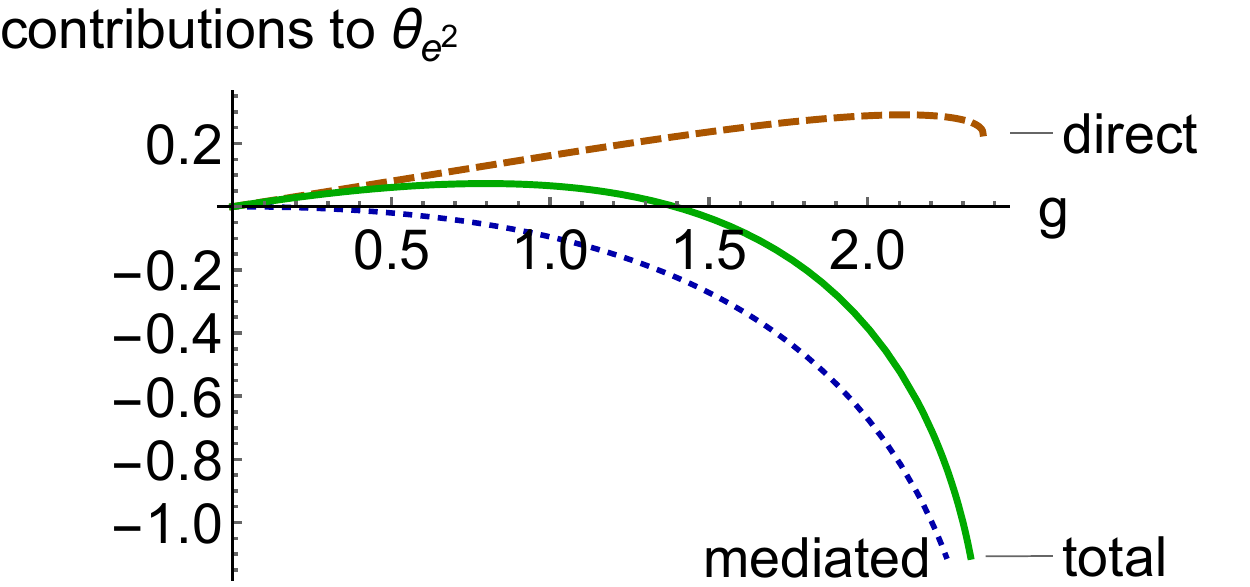}
\caption{\label{fig:theta_alpha} We show the two competing contributions to 
$\theta_{e^2}$ from eq.~\eqref{eq:thetaalphag} and 
eq.~\eqref{eq:thetaalphaw2} as well as the full result, both for $\eta_h=0$, with eq.~\eqref{eq:etaAfull}. At 
$g= g_{\rm zero} \approx 1.5$, the gauge coupling $e$ becomes irrelevant, 
making a connection of the fixed point to phenomenologically acceptable values 
difficult. 
 $g_{\mathrm{zero}}$ gets 
smaller for $\mu_h <0$, while the influence of $\eta_h$ is negligible.}
\end{figure}
The shifted Gau\ss{}ian fixed point  in $w_2$ evolves towards 
 increasingly negative
values with
 growing
gravitational coupling $g$,  growing quadratically in $g$ for small $g$. Thus, the contribution to 
$\theta_{e^2}$ from the induced gauge coupling $w_2$ ultimately wins over 
the direct quantum gravity contribution, cf.~Fig.~\ref{fig:theta_alpha}.\\
 For small gravitational coupling, we thus have a fixed point at which the 
dimensionless gauge coupling $e^2$ is asymptotically free, while $w_2$ becomes 
asymptotically safe and corresponds to an irrelevant direction of the fixed 
point. This is the phenomenologically preferred regime for $g$, as it provides a 
UV completion for the Abelian gauge sector  -- thus solving the triviality 
problem -- that can be matched onto the high-energy behavior of the hypercharge 
gauge coupling in the Standard Model.  Inserting fixed-point values for $g, 
\mu_h$ and $\eta_h$ from the truncation explored in \cite{Dona:2013qba}, we find 
$\theta_{e^2}\approx 0.07$, i.e., indications point towards a 
phenomenologically viable quantum-gravity induced solution of the triviality 
problem.\\
Towards larger $g$, 
the gauge coupling  $e$ is then rendered irrelevant at its Gau\ss{}ian 
fixed point, before the full fixed point for the system is destroyed at $g 
=g_{\rm crit}$. In the intermediate regime, i.e., between $g=g_{\rm zero} 
\approx 1.5$ and $g< g_{\rm crit} \approx 2.4$ (all values for $\mu_h=0, \, 
\eta_h=0$), all couplings in the gauge sector feature fixed points and are 
irrelevant,
 i.e.,
the fixed point is completely UV-repulsive, i.e., completely 
IR-attractive. This means that quantum-gravity 
fluctuations force those couplings to remain at their fixed-point values all the 
way down to the Planck scale and $e(k \approx M_{\mathrm{Pl}})=0$. At the 
Planck 
scale, quantum-gravity effects switch off, and the further running of the 
couplings down to the electroweak scale is determined purely in terms of 
Standard Model beta functions.
In that regime, the low-energy values of those couplings therefore correspond to 
a prediction of the asymptotic safety paradigm, see \cite{Shaposhnikov:2009pv} for a similar argument in the Higgs sector. In a setting where no new 
physics appears between the electroweak and the Planck scale, the U(1) 
hypercharge coupling is significantly larger than 0 at the 
Planck scale. However, this value cannot be reached starting 
from a fixed point at $e_*=0$, if that coupling is irrelevant and thus 
forced to remain zero until quantum gravity fluctuations switch off.

If these results persist beyond our truncation, and the 
assumption of no new physics up to the Planck scale holds, we would tentatively 
conclude that the gravitational coupling should not become too large in the UV, 
such that $e$ can in fact remain asymptotically free. Interestingly, the 
excluded region in Fig.~\ref{fig:exclusions} is similar to that found in 
\cite{Eichhorn:2016esv}, where quantum-gravity fluctuations push an interacting 
fixed point in the Yukawa sector into the complex plane. We thus observe hints 
from two separate sectors of the Standard Model, that quantum-gravity 
fluctuations are restricted to not exceed a critical strength, if the asymptotic 
safety paradigm is to provide a viable UV completion for gravity and matter.

Broadening our scope beyond the asymptotic-safety scenario, we observe that the generation of higher-order operators by quantum gravity appears to be a generic feature within the effective field theory setting \cite{EFT}: Any non-zero value of $g$ provides a contribution $\sim g^2$ to the beta function for $w_2$, i.e., even if $w_2$ is set to zero at some scale, quantum gravity fluctuations generate it. Our results suggest that going beyond canonical power counting might also be important to understand the full effect of quantum gravity on gauge couplings in the effective field theory framework \cite{EFTabelian}.

In particular, an analysis within the EFT regime for gravity applies to the case where asymptotically safe matter models \cite{AS4d} are subjected to quantum-gravity corrections. In that case, corrections linear in the gauge coupling as in eq.~\eqref{eq:betaalpha_simp} arise, and contribute towards asymptotic freedom in the gauge coupling,  potentially impacting the phase structure in those models, \cite{Christiansen2017}.

\section{Conclusions and outlook}
In this work, we have found evidence that asymptotically safe quantum gravity 
can generate an interacting fixed point for a U(1) gauge theory, thus providing 
a solution to the triviality problem. We have further elucidated the fixed-point 
structure by pointing out that it is necessary to go beyond canonically 
dimensionless couplings in the gauge sector: Asymptotically safe quantum gravity 
induces non-vanishing higher order interactions. Therefore, it is impossible for 
an Abelian gauge theory to become completely asymptotically free under the 
impact of quantum gravity: While the leading-order gauge coupling $e$ 
features a fixed point at zero, the higher-order coupling $w_2$ is 
non-vanishing. This effect becomes of critical importance for two questions: 
Firstly, the very existence of a quantum-gravity generated fixed point in the 
gauge sector looks very different when higher-order couplings are neglected. 
This is the case since quantum gravity first shifts the Gau\ss{}ian fixed point 
for the higher order couplings to non-zero real values and then to complex 
values once a critical strength of gravitational interactions is exceeded. 
Secondly, the inclusion of the higher-order coupling $w_2$ is  critical to 
understand whether $e$ can become asymptotically free under the impact of 
quantum gravity: Once generated by quantum gravity, $w_2$ yields a nonzero 
contribution to the scaling dimension of $e$ at its free fixed point. 
Beyond a critical strength of the gravitational coupling, the gauge coupling is 
then rendered irrelevant at its fixed point. This setting  
 is
difficult 
to reconcile with the Planck-scale value of the hypercharge coupling  unless additional new physics is invoked.
\\
We  tentatively conclude that quantum gravity can provide a
UV completion for Abelian gauge theories within the asymptotic-safety paradigm. 
 Crucially,
this mechanism relies on the strength of gravitational interactions not exceeding a certain critical value. Beyond the shifted Gau\ss{}ian fixed point, further fully 
 interacting
fixed points could exist, but these would typically  feature significantly different sets of relevant operators.
Interestingly, quantum fluctuations of minimally coupled gauge fields force the fixed-point value for the Newton coupling towards lower values, see \cite{Dona:2013qba}. Therefore, even if gravity destroys the fixed point for one gauge field, the shift towards weaker gravity that is induced by coupling additional gauge fields 
 will
ultimately shift the fixed point for $w_2$ back onto the real axis.
Thus, a scenario could be realized, in which a gauge-gravity system can only be asymptotically safe if a critical lower number of gauge fields is exceeded.
With the caveat that these were obtained with a different regularization scheme and pertain to the background field, we can insert the fixed-point values for $g$, $\eta_h$ and $\mu$ from \cite{Dona:2013qba} for one vector field, and we find that the shifted Gau\ss{}ian fixed point already exists. Studies of (non-minimally) gauge-gravity systems will have to confirm this result. 
 
Our work paves the way for 
 several
exciting questions to be explored in the future: Firstly, similar mechanisms will exist in models with additional (``dark") U(1) sectors, and will induce couplings between photons and hidden photons. It will be interesting to explore the properties of an asymptotically safe portal to the dark sector.\\
Moreover, our results also carry over to the case of non-Abelian gauge theories, 
where similar higher-order operators will be induced. It is critical to test 
their impact on the running of the non-Abelian gauge coupling, as quantum 
gravity could then either destroy asymptotic freedom, strengthen it, or turn it 
into asymptotic safety 
\cite{Daum:2009dn,Folkerts:2011jz,ChristiansenLitimPawlowski2017}. In that case, 
the non-Abelian gauge couplings could even be become irrelevant couplings, and 
their low-energy values would be a prediction of asymptotic safety.\\
Further, it is known that non-Abelian gauge theories are asymptotically safe by themselves in $d=4+\epsilon$ dimensions \cite{Peskin:1980ay,Gies:2003ic}. 
 Thus, a setting in $d>4$ could feature an intricate interplay of fixed-point dynamics for the gauge and the gravity sector.

Moreover, our result that quantum gravity induces higher-order interactions in gauge theories, which then affect the question whether the gauge theory can become asymptotically free under the effect of quantum gravity, is not restricted to asymptotically safe gravity, as it does not require a particular value for the Newton coupling. Thus, our work suggests that going beyond canonical power counting could be necessary to obtain a full picture of the effect of quantum gravity also within an effective field theory setting. In particular, it would be interesting to explore how the effects that we have observed here are encoded in other schemes. 

\emph{Acknowledgements:}
We acknowledge helpful discussions with A.~Held.
This work was supported by the DFG under the Emmy-Noether program, grant  no.~EI-1037-1. A.~E.~is also supported by an Emmy-Noether visiting fellowship at the Perimeter Institute for Theoretical Physics. Research at Perimeter Institute is supported by the Government of Canada
through Industry Canada and by the Province of Ontario through the Ministry of Economic Development \&
Innovation.


\begin{thebibliography}{99}
%\cite{Weinberg:1980gg}
\bibitem{Weinberg:1980gg}
  S.~Weinberg,
%  ``Ultraviolet Divergences In Quantum Theories Of Gravitation,''
%\href{http://www.slac.stanford.edu/spires/find/hep/www?irn=784877}{SPIRES entry}
{\it  In *Hawking, S.W., Israel, W.: General Relativity*, 790-831}
(Cambridge University Press, Cambridge, 1980).


%
\bibitem{GellMann:1954fq} 
  M.~Gell-Mann and F.~E.~Low,
  %``Quantum electrodynamics at small distances,''
  Phys.\ Rev.\  {\bf 95}, 1300 (1954).
  doi:10.1103/PhysRev.95.1300
  %%CITATION = doi:10.1103/PhysRev.95.1300;%%
  
  %\cite{Gockeler:1997dn}
\bibitem{Gockeler:1997dn} 
  M.~Gockeler, R.~Horsley, V.~Linke, P.~E.~L.~Rakow, G.~Schierholz and H.~Stuben,
  %``Is there a Landau pole problem in QED?,''
  Phys.\ Rev.\ Lett.\  {\bf 80}, 4119 (1998)
  doi:10.1103/PhysRevLett.80.4119
  [hep-th/9712244].
  %%CITATION = doi:10.1103/PhysRevLett.80.4119;%%
  
  \bibitem{Gockeler:1997kt} 
  M.~Gockeler, R.~Horsley, V.~Linke, P.~E.~L.~Rakow, G.~Schierholz and H.~Stuben,
  %``Resolution of the Landau pole problem in QED,''
  Nucl.\ Phys.\ Proc.\ Suppl.\  {\bf 63}, 694 (1998)
  doi:10.1016/S0920-5632(97)00875-X
  [hep-lat/9801004].
  %%CITATION = doi:10.1016/S0920-5632(97)00875-X;%%
  
%\cite{Gies:2004hy}
\bibitem{Gies:2004hy} 
  H.~Gies and J.~Jaeckel,
  %``Renormalization flow of QED,''
  Phys.\ Rev.\ Lett.\  {\bf 93}, 110405 (2004)
  doi:10.1103/PhysRevLett.93.110405
  [hep-ph/0405183].
  %%CITATION = doi:10.1103/PhysRevLett.93.110405;%%


%\cite{Harst:2011zx}
\bibitem{Harst:2011zx} 
  U.~Harst and M.~Reuter,
  %``QED coupled to QEG,''
  JHEP {\bf 1105}, 119 (2011)
  doi:10.1007/JHEP05(2011)119
  [arXiv:1101.6007 [hep-th]].
  %%CITATION = doi:10.1007/JHEP05(2011)119;%%

  %\cite{Eichhorn:2011pc}
\bibitem{Eichhorn:2011pc} 
  A.~Eichhorn and H.~Gies,
  %``Light fermions in quantum gravity,''
  New J.\ Phys.\  {\bf 13}, 125012 (2011)
  doi:10.1088/1367-2630/13/12/125012
  [arXiv:1104.5366 [hep-th]].
  %%CITATION = doi:10.1088/1367-2630/13/12/125012;%%

%\cite{Eichhorn:2011ec}
\bibitem{Eichhorn:2011ec} 
  A.~Eichhorn,
  %``Observable consequences of quantum gravity: Can light fermions exist?,''
  J.\ Phys.\ Conf.\ Ser.\  {\bf 360}, 012057 (2012)
  doi:10.1088/1742-6596/360/1/012057
  [arXiv:1109.3784 [gr-qc]].
  %%CITATION = doi:10.1088/1742-6596/360/1/012057;%%
  
    %\cite{Eichhorn:2012va}
\bibitem{Eichhorn:2012va} 
  A.~Eichhorn,
  %``Quantum-gravity-induced matter self-interactions in the asymptotic-safety scenario,''
  Phys.\ Rev.\ D {\bf 86}, 105021 (2012)
  doi:10.1103/PhysRevD.86.105021
  [arXiv:1204.0965 [gr-qc]].
  %%CITATION = doi:10.1103/PhysRevD.86.105021;%%
  
  %\cite{Eichhorn:2013ug}
\bibitem{Eichhorn:2013ug} 
  A.~Eichhorn,
  %``Faddeev-Popov ghosts in quantum gravity beyond perturbation theory,''
  Phys.\ Rev.\ D {\bf 87}, no. 12, 124016 (2013)
  doi:10.1103/PhysRevD.87.124016
  [arXiv:1301.0632 [hep-th]].
  %%CITATION = doi:10.1103/PhysRevD.87.124016;%%

  
    %\cite{Meibohm:2016mkp}
\bibitem{Meibohm:2016mkp} 
  J.~Meibohm and J.~M.~Pawlowski,
  %``Chiral fermions in asymptotically safe quantum gravity,''
  Eur.\ Phys.\ J.\ C {\bf 76}, no. 5, 285 (2016)
  doi:10.1140/epjc/s10052-016-4132-7
  [arXiv:1601.04597 [hep-th]].
  %%CITATION = doi:10.1140/epjc/s10052-016-4132-7;%%
  
  \bibitem{Eichhorn:2016esv} 
  A.~Eichhorn, A.~Held and J.~M.~Pawlowski,
  %``Quantum-gravity effects on a Higgs-Yukawa model,''
  Phys.\ Rev.\ D {\bf 94}, no. 10, 104027 (2016)
  doi:10.1103/PhysRevD.94.104027
  [arXiv:1604.02041 [hep-th]].
  %%CITATION = doi:10.1103/PhysRevD.94.104027;%%
  

 %\cite{Eichhorn:2017eht}
\bibitem{Eichhorn:2017eht} 
  A.~Eichhorn and A.~Held,
  %``Viability of quantum-gravity induced ultraviolet completions for matter,''
  arXiv:1705.02342 [gr-qc].
  %%CITATION = ARXIV:1705.02342;%%


  
  %\cite{Litim:2001up}
\bibitem{Litim:2001up} 
  D.~F.~Litim,
  %``Optimized renormalization group flows,''
  Phys.\ Rev.\ D {\bf 64}, 105007 (2001)
  doi:10.1103/PhysRevD.64.105007
  [hep-th/0103195].
  %%CITATION = doi:10.1103/PhysRevD.64.105007;%%
  
      %\cite{Wetterich:1992yh}
\bibitem{Wetterich:1992yh} 
  C.~Wetterich,
  %``Exact evolution equation for the effective potential,''
  Phys.\ Lett.\ B {\bf 301}, 90 (1993).
%   doi:10.1016/0370-2693(93)90726-X
  %%CITATION = doi:10.1016/0370-2693(93)90726-X;%%
  
        %
\bibitem{Morris:1993qb} 
  T.~R.~Morris,
  %``The Exact renormalization group and approximate solutions,''
  Int.\ J.\ Mod.\ Phys.\ A {\bf 9}, 2411 (1994)
%  doi:10.1142/S0217751X94000972
  [hep-ph/9308265].
  %%CITATION = doi:10.1142/S0217751X94000972;%%
  
  %\cite{Ellwanger:1993mw}
\bibitem{Ellwanger:1993mw} 
  U.~Ellwanger,
  %``FLow equations for N point functions and bound states,''
  Z.\ Phys.\ C {\bf 62}, 503 (1994)
%   doi:10.1007/BF01555911
  [hep-ph/9308260].
  %%CITATION = doi:10.1007/BF01555911;%%
  
   \bibitem{Berges:2000ew}
  J.~Berges, N.~Tetradis and C.~Wetterich,
  Phys.\ Rept.\  {\bf 363} (2002) 223
  [hep-ph/0005122].
  %%CITATION = HEP-PH 0005122;%%
  
  \bibitem{Polonyi:2001se}
  J.~Polonyi,
  %``Lectures on the functional renormalization group
%method,''
  Central Eur.\ J.\ Phys.\  {\bf 1}, 1 (2003) 
 [hep-th/0110026].
  %%CITATION = HEP-TH 0110026;%%
%
%\cite{Pawlowski:2005xe}
\bibitem{Pawlowski:2005xe}
  J.~M.~Pawlowski,
 % ``Aspects of the functional renormalisation group,''
  Annals Phys.\  {\bf 322} (2007) 2831 
  [arXiv:hep-th/0512261].
  %%CITATION = APNYA,322,2831;%%
%

%\cite{Gies:2006wv}
\bibitem{Gies:2006wv}
  H.~Gies, Lect.\ Notes Phys.\ {\bf 852}, 287 (2012)
  %``Introduction to the functional RG and applications to gauge theories,''
  [arXiv:hep-ph/0611146]. %.
  %%CITATION = HEP-PH/0611146;%%

%\cite{Delamotte:2007pf}
\bibitem{Delamotte:2007pf} 
  B.~Delamotte,
  %``An Introduction to the nonperturbative renormalization group,''
  Lect.\ Notes Phys.\  {\bf 852}, 49 (2012)
  [cond-mat/0702365].
  %%CITATION = COND-MAT/0702365;%%
  
%\cite{Rosten:2010vm}
\bibitem{Rosten:2010vm}
  O.~J.~Rosten,
%  ``Fundamentals of the Exact Renormalization Group,''
  arXiv:1003.1366 [hep-th].
  %%CITATION = ARXIV:1003.1366;%%
  
  %\cite{Braun:2011pp}
\bibitem{Braun:2011pp} 
  J.~Braun,
  %``Fermion Interactions and Universal Behavior in Strongly Interacting Theories,''
  J.\ Phys.\ G {\bf 39}, 033001 (2012)
%   doi:10.1088/0954-3899/39/3/033001
  [arXiv:1108.4449 [hep-ph]].
  %%CITATION = doi:10.1088/0954-3899/39/3/033001;%%
  
    %
\bibitem{Reuter:1996cp} 
  M.~Reuter,
  %``non-perturbative evolution equation for quantum gravity,''
  Phys.\ Rev.\ D {\bf 57}, 971 (1998)
  [hep-th/9605030].
  %%CITATION = HEP-TH/9605030;%%
  
  %\cite{Reuter:2001ag}
\bibitem{Reuter:2001ag} 
  M.~Reuter and F.~Saueressig,
  %``Renormalization group flow of quantum gravity in the Einstein-Hilbert truncation,''
  Phys.\ Rev.\ D {\bf 65}, 065016 (2002)
  [hep-th/0110054].
  %%CITATION = HEP-TH/0110054;%%
  
  \bibitem{Lauscher:2001ya} 
  O.~Lauscher and M.~Reuter,
  %``Ultraviolet fixed point and generalized flow equation of quantum gravity,''
  Phys.\ Rev.\ D {\bf 65}, 025013 (2002)
  doi:10.1103/PhysRevD.65.025013
  [hep-th/0108040].
  %%CITATION = doi:10.1103/PhysRevD.65.025013;%%
  

%\cite{Lauscher:2002sq}
\bibitem{Lauscher:2002sq} 
  O.~Lauscher and M.~Reuter,
  %``Flow equation of quantum Einstein gravity in a higher derivative truncation,''
  Phys.\ Rev.\ D {\bf 66}, 025026 (2002)
  [hep-th/0205062].
  %%CITATION = HEP-TH/0205062;%%

\bibitem{Litim:2003vp} 
  D.~F.~Litim,
  %``Fixed points of quantum gravity,''
  Phys.\ Rev.\ Lett.\  {\bf 92}, 201301 (2004)
  [hep-th/0312114].
  %%CITATION = HEP-TH/0312114;%%

  
  %\cite{Fischer:2006fz}
\bibitem{Fischer:2006fz} 
  P.~Fischer and D.~F.~Litim,
  %``Fixed points of quantum gravity in extra dimensions,''
  Phys.\ Lett.\ B {\bf 638}, 497 (2006)
  [hep-th/0602203].
  %%CITATION = HEP-TH/0602203;%%

%\cite{Machado:2007ea}
\bibitem{Machado:2007ea} 
  P.~F.~Machado and F.~Saueressig,
  %``On the renormalization group flow of f(R)-gravity,''
  Phys.\ Rev.\ D {\bf 77}, 124045 (2008)
  [arXiv:0712.0445 [hep-th]].
  %%CITATION = ARXIV:0712.0445;%%

%\cite{}
\bibitem{Codello:2006in} 
  A.~Codello and R.~Percacci,
  %``Fixed points of higher derivative gravity,''
  Phys.\ Rev.\ Lett.\  {\bf 97}, 221301 (2006)
  [hep-th/0607128].
  %%CITATION = HEP-TH/0607128;%%
  
  %
\bibitem{Codello:2008vh} 
  A.~Codello, R.~Percacci and C.~Rahmede,
  %``Investigating the Ultraviolet Properties of Gravity with a Wilsonian Renormalization Group Equation,''
  Annals Phys.\  {\bf 324}, 414 (2009)
  [arXiv:0805.2909 [hep-th]].
  %%CITATION = ARXIV:0805.2909;%%
    
%\cite{Eichhorn:2009ah}
\bibitem{Eichhorn:2009ah} 
  A.~Eichhorn, H.~Gies and M.~M.~Scherer,
  %``Asymptotically free scalar curvature-ghost coupling in Quantum Einstein Gravity,''
  Phys.\ Rev.\ D {\bf 80}, 104003 (2009)
  [arXiv:0907.1828 [hep-th]].
  %%CITATION = ARXIV:0907.1828;%%
  
  %\cite{Manrique:2009uh}
\bibitem{Manrique:2009uh} 
  E.~Manrique and M.~Reuter,
  %``Bimetric Truncations for Quantum Einstein Gravity and Asymptotic Safety,''
  Annals Phys.\  {\bf 325}, 785 (2010)
  doi:10.1016/j.aop.2009.11.009
  [arXiv:0907.2617 [gr-qc]].
  %%CITATION = doi:10.1016/j.aop.2009.11.009;%%
  
%\cite{Benedetti:2009rx}
\bibitem{Benedetti:2009rx} 
  D.~Benedetti, P.~F.~Machado and F.~Saueressig,
  %``Asymptotic safety in higher-derivative gravity,''
  Mod.\ Phys.\ Lett.\ A {\bf 24}, 2233 (2009)
  [arXiv:0901.2984 [hep-th]].
  %%CITATION = ARXIV:0901.2984;%%
  
  %\cite{Eichhorn:2010tb}
\bibitem{Eichhorn:2010tb} 
  A.~Eichhorn and H.~Gies,
  %``Ghost anomalous dimension in asymptotically safe quantum gravity,''
  Phys.\ Rev.\ D {\bf 81}, 104010 (2010)
  [arXiv:1001.5033 [hep-th]].
  %%CITATION = ARXIV:1001.5033;%%

%\cite{Groh:2010ta}
\bibitem{Groh:2010ta} 
  K.~Groh and F.~Saueressig,
  %``Ghost wave-function renormalization in Asymptotically Safe Quantum Gravity,''
  J.\ Phys.\ A {\bf 43}, 365403 (2010)
  [arXiv:1001.5032 [hep-th]].
  %%CITATION = ARXIV:1001.5032;%%
  
  %\cite{Manrique:2010mq}
\bibitem{Manrique:2010mq} 
  E.~Manrique, M.~Reuter and F.~Saueressig,
  %``Matter Induced Bimetric Actions for Gravity,''
  Annals Phys.\  {\bf 326}, 440 (2011)
  doi:10.1016/j.aop.2010.11.003
  [arXiv:1003.5129 [hep-th]].
  %%CITATION = doi:10.1016/j.aop.2010.11.003;%%
  
  %\cite{Manrique:2010am}
\bibitem{Manrique:2010am} 
  E.~Manrique, M.~Reuter and F.~Saueressig,
  %``Bimetric Renormalization Group Flows in Quantum Einstein Gravity,''
  Annals Phys.\  {\bf 326}, 463 (2011)
  doi:10.1016/j.aop.2010.11.006
  [arXiv:1006.0099 [hep-th]].
  %%CITATION = doi:10.1016/j.aop.2010.11.006;%%
  
  %\cite{Manrique:2011jc}
\bibitem{Manrique:2011jc} 
  E.~Manrique, S.~Rechenberger and F.~Saueressig,
  %``Asymptotically Safe Lorentzian Gravity,''
  Phys.\ Rev.\ Lett.\  {\bf 106}, 251302 (2011)
  [arXiv:1102.5012 [hep-th]].
  %%CITATION = ARXIV:1102.5012;%%
  
  %\cite{Benedetti:2012dx}
\bibitem{Benedetti:2012dx} 
  D.~Benedetti and F.~Caravelli,
  %``The Local potential approximation in quantum gravity,''
  JHEP {\bf 1206}, 017 (2012)
  [Erratum-ibid.\  {\bf 1210}, 157 (2012)]
  [arXiv:1204.3541 [hep-th]].
  %%CITATION = ARXIV:1204.3541;%

%\cite{Christiansen:2012rx}
\bibitem{Christiansen:2012rx} 
  N.~Christiansen, D.~F.~Litim, J.~M.~Pawlowski and A.~Rodigast,
  %``Fixed points and infrared completion of quantum gravity,''
  Phys.\ Lett.\ B {\bf 728}, 114 (2014)
  [arXiv:1209.4038 [hep-th]].
  %%CITATION = ARXIV:1209.4038;%%
  
%\cite{Dietz:2012ic}
\bibitem{Dietz:2012ic} 
  J.~A.~Dietz and T.~R.~Morris,
 %``Asymptotic safety in the f(R) approximation,''
  JHEP {\bf 1301}, 108 (2013)
  doi:10.1007/JHEP01(2013)108
  [arXiv:1211.0955 [hep-th]].
  %%CITATION = doi:10.1007/JHEP01(2013)108;%%
        
    %\cite{}
\bibitem{Falls:2013bv} 
  K.~Falls, D.~F.~Litim, K.~Nikolakopoulos and C.~Rahmede,
  %``A bootstrap towards asymptotic safety,''
  arXiv:1301.4191 [hep-th].
  %%CITATION = ARXIV:1301.4191;%%
  
%\cite{Christiansen:2014raa}
\bibitem{Christiansen:2014raa} 
  N.~Christiansen, B.~Knorr, J.~M.~Pawlowski and A.~Rodigast,
  %``Global Flows in Quantum Gravity,''
  Phys.\ Rev.\ D {\bf 93}, no. 4, 044036 (2016)
  doi:10.1103/PhysRevD.93.044036
  [arXiv:1403.1232 [hep-th]].
  %%CITATION = doi:10.1103/PhysRevD.93.044036;%%
  
   \bibitem{Becker:2014qya} 
  D.~Becker and M.~Reuter,
  %``En route to Background Independence: Broken split-symmetry, and how to restore it with bi-metric average actions,''
  Annals Phys.\  {\bf 350}, 225 (2014)
  doi:10.1016/j.aop.2014.07.023
  [arXiv:1404.4537 [hep-th]].
  %%CITATION = doi:10.1016/j.aop.2014.07.023;%%
         
  %\cite{Falls:2014tra}
\bibitem{Falls:2014tra} 
  K.~Falls, D.~F.~Litim, K.~Nikolakopoulos and C.~Rahmede,
  %``Further evidence for asymptotic safety of quantum gravity,''
  Phys.\ Rev.\ D {\bf 93}, no. 10, 104022 (2016)
  doi:10.1103/PhysRevD.93.104022
  [arXiv:1410.4815 [hep-th]].
  %%CITATION = doi:10.1103/PhysRevD.93.104022;%%
  
   %\cite{Eichhorn:2015bna}
\bibitem{Eichhorn:2015bna} 
  A.~Eichhorn,
  %``The Renormalization Group flow of unimodular f(R) gravity,''
  JHEP {\bf 1504}, 096 (2015)
  [arXiv:1501.05848 [gr-qc]].
  %%CITATION = ARXIV:1501.05848;%%
  
  \bibitem{Demmel:2015oqa} 
  M.~Demmel, F.~Saueressig and O.~Zanusso,
  %``A proper fixed functional for four-dimensional Quantum Einstein Gravity,''
  JHEP {\bf 1508}, 113 (2015)
  doi:10.1007/JHEP08(2015)113
  [arXiv:1504.07656 [hep-th]].
  %%CITATION = doi:10.1007/JHEP08(2015)113;%%
  
%\cite{Christiansen:2015rva}
\bibitem{Christiansen:2015rva} 
  N.~Christiansen, B.~Knorr, J.~Meibohm, J.~M.~Pawlowski and M.~Reichert,
  %``Local Quantum Gravity,''
  Phys.\ Rev.\ D {\bf 92}, no. 12, 121501 (2015)
  doi:10.1103/PhysRevD.92.121501
  [arXiv:1506.07016 [hep-th]].
  %%CITATION = doi:10.1103/PhysRevD.92.121501;%%
  
%\cite{Gies:2015tca}
\bibitem{Gies:2015tca} 
  H.~Gies, B.~Knorr and S.~Lippoldt,
  %``Generalized Parametrization Dependence in Quantum Gravity,''
  Phys.\ Rev.\ D {\bf 92}, no. 8, 084020 (2015)
  doi:10.1103/PhysRevD.92.084020
  [arXiv:1507.08859 [hep-th]].
  %%CITATION = doi:10.1103/PhysRevD.92.084020;%%
  
  %\cite{}
\bibitem{Morris:2016spn} 
  T.~R.~Morris,
  %``Large curvature and background scale independence in single-metric approximations to asymptotic safety,''
  JHEP {\bf 1611}, 160 (2016)
  doi:10.1007/JHEP11(2016)160
  [arXiv:1610.03081 [hep-th]].
  %%CITATION = doi:10.1007/JHEP11(2016)160;%%
  
  %\cite{Percacci:2016arh}
\bibitem{Percacci:2016arh} 
  R.~Percacci and G.~P.~Vacca,
  %``The background scale Ward identity in quantum gravity,''
  Eur.\ Phys.\ J.\ C {\bf 77}, no. 1, 52 (2017)
  doi:10.1140/epjc/s10052-017-4619-x
  [arXiv:1611.07005 [hep-th]].
  %%CITATION = doi:10.1140/epjc/s10052-017-4619-x;%%
  
  %\cite{Gies:2016con}
\bibitem{Gies:2016con} 
  H.~Gies, B.~Knorr, S.~Lippoldt and F.~Saueressig,
  %``Gravitational Two-Loop Counterterm Is Asymptotically Safe,''
  Phys.\ Rev.\ Lett.\  {\bf 116}, no. 21, 211302 (2016)
  doi:10.1103/PhysRevLett.116.211302
  [arXiv:1601.01800 [hep-th]].
  %%CITATION = doi:10.1103/PhysRevLett.116.211302;%%
  
  %\cite{Ohta:2016npm}
\bibitem{Ohta:2016npm} 
  N.~Ohta, R.~Percacci and A.~D.~Pereira,
  %``Gauges and functional measures in quantum gravity I: Einstein theory,''
  JHEP {\bf 1606}, 115 (2016)
  doi:10.1007/JHEP06(2016)115
  [arXiv:1605.00454 [hep-th]].
  %%CITATION = doi:10.1007/JHEP06(2016)115;%%
  
  %\cite{Henz:2016aoh}
\bibitem{Henz:2016aoh} 
  T.~Henz, J.~M.~Pawlowski and C.~Wetterich,
  %``Scaling solutions for Dilaton Quantum Gravity,''
  doi:10.1016/j.physletb.2017.01.057
  arXiv:1605.01858 [hep-th].
  %%CITATION = doi:10.1016/j.physletb.2017.01.057;%%
   
   %\cite{Biemans:2016rvp}
\bibitem{Biemans:2016rvp} 
  J.~Biemans, A.~Platania and F.~Saueressig,
  %``Quantum gravity on foliated spacetime - asymptotically safe and sound,''
  arXiv:1609.04813 [hep-th].
  %%CITATION = ARXIV:1609.04813;%%
  
   %\cite{Pagani:2016dof}
\bibitem{Pagani:2016dof} 
  C.~Pagani and M.~Reuter,
  %``Composite Operators in Asymptotic Safety,''
  arXiv:1611.06522 [gr-qc].
  %%CITATION = ARXIV:1611.06522;%%
  
     %\cite{Christiansen:2016sjn}
\bibitem{Christiansen:2016sjn} 
  N.~Christiansen,
  %``Four-Derivative Quantum Gravity Beyond Perturbation Theory,''
  arXiv:1612.06223 [hep-th].
  %%CITATION = ARXIV:1612.06223;%%
  
  %\cite{Denz:2016qks}
\bibitem{Denz:2016qks} 
  T.~Denz, J.~M.~Pawlowski and M.~Reichert,
  %``Towards apparent convergence in asymptotically safe quantum gravity,''
  arXiv:1612.07315 [hep-th].
  %%CITATION = ARXIV:1612.07315;%%
  
  %\cite{Ohta:2017dsq}
\bibitem{Ohta:2017dsq} 
  N.~Ohta,
  %``Background Scale Independence in Quantum Gravity,''
  arXiv:1701.01506 [hep-th].
  %%CITATION = ARXIV:1701.01506;%%
   
%\cite{Falls:2017cze}
\bibitem{Falls:2017cze} 
  K.~Falls,
  %``Physical renormalisation schemes and asymptotic safety in quantum gravity,''
  arXiv:1702.03577 [hep-th].
  %%CITATION = ARXIV:1702.03577;%%
 
 %\cite{Benedetti:2009gn}
\bibitem{Benedetti:2009gn} 
  D.~Benedetti, P.~F.~Machado and F.~Saueressig,
  %``Taming perturbative divergences in asymptotically safe gravity,''
  Nucl.\ Phys.\ B {\bf 824}, 168 (2010)
  doi:10.1016/j.nuclphysb.2009.08.023
  [arXiv:0902.4630 [hep-th]].
  %%CITATION = doi:10.1016/j.nuclphysb.2009.08.023;%%
  
  %\cite{Dona:2012am}
\bibitem{Dona:2012am} 
  P.~Don\`a and R.~Percacci,
  %``Functional renormalization with fermions and tetrads,''
  Phys.\ Rev.\ D {\bf 87}, no. 4, 045002 (2013)
  doi:10.1103/PhysRevD.87.045002
  [arXiv:1209.3649 [hep-th]].
  %%CITATION = doi:10.1103/PhysRevD.87.045002;%%    
          
    %\cite{Dona:2013qba}
\bibitem{Dona:2013qba} 
  P.~Don\`a, A.~Eichhorn and R.~Percacci,
  %``Matter matters in asymptotically safe quantum gravity,''
  Phys.\ Rev.\ D {\bf 89}, no. 8, 084035 (2014)
  doi:10.1103/PhysRevD.89.084035
  [arXiv:1311.2898 [hep-th]].
  %%CITATION = doi:10.1103/PhysRevD.89.084035;%%
  
    %\cite{Dona:2014pla}
\bibitem{Dona:2014pla} 
  P.~Don\`a, A.~Eichhorn and R.~Percacci,
  %``Consistency of matter models with asymptotically safe quantum gravity,''
  Can.\ J.\ Phys.\  {\bf 93}, no. 9, 988 (2015)
  doi:10.1139/cjp-2014-0574
  [arXiv:1410.4411 [gr-qc]].
  %%CITATION = doi:10.1139/cjp-2014-0574;%%
  
    %\cite{Oda:2015sma}
\bibitem{Oda:2015sma} 
  K.~y.~Oda and M.~Yamada,
  %``Non-minimal coupling in Higgs?Yukawa model with asymptotically safe gravity,''
  Class.\ Quant.\ Grav.\  {\bf 33}, no. 12, 125011 (2016)
  doi:10.1088/0264-9381/33/12/125011
  [arXiv:1510.03734 [hep-th]].
  %%CITATION = doi:10.1088/0264-9381/33/12/125011;%%
  
    %\cite{Meibohm:2015twa}
\bibitem{Meibohm:2015twa} 
  J.~Meibohm, J.~M.~Pawlowski and M.~Reichert,
  %``Asymptotic safety of gravity-matter systems,''
  Phys.\ Rev.\ D {\bf 93}, no. 8, 084035 (2016)
  doi:10.1103/PhysRevD.93.084035
  [arXiv:1510.07018 [hep-th]].
  %%CITATION = doi:10.1103/PhysRevD.93.084035;%%
  
  %\cite{Dona:2015tnf}
\bibitem{Dona:2015tnf} 
  P.~Don\`a, A.~Eichhorn, P.~Labus and R.~Percacci,
  %``Asymptotic safety in an interacting system of gravity and scalar matter,''
  Phys.\ Rev.\ D {\bf 93}, no. 4, 044049 (2016)
  Erratum: [Phys.\ Rev.\ D {\bf 93}, no. 12, 129904 (2016)]
  doi:10.1103/PhysRevD.93.129904, 10.1103/PhysRevD.93.044049
  [arXiv:1512.01589 [gr-qc]].
  %%CITATION = doi:10.1103/PhysRevD.93.129904, 10.1103/PhysRevD.93.044049;%%
  
%\cite{Biemans:2017zca}
\bibitem{Biemans:2017zca} 
  J.~Biemans, A.~Platania and F.~Saueressig,
  %``Renormalization group fixed points of foliated gravity-matter systems,''
  arXiv:1702.06539 [hep-th].
  %%CITATION = ARXIV:1702.06539;%%
  
   \bibitem{ASreviews}
  %\cite{Niedermaier:2006wt}
%\bibitem{Niedermaier:2006wt} 
  M.~Niedermaier and M.~Reuter,
  %``The Asymptotic Safety Scenario in Quantum Gravity,''
  Living Rev.\ Rel.\  {\bf 9}, 5 (2006);
  %%CITATION = 00222,9,5;%%
%\cite{}
%\bibitem{Niedermaier:2006ns} 
  M.~Niedermaier,
  %``The Asymptotic safety scenario in quantum gravity: An Introduction,''
  Class.\ Quant.\ Grav.\  {\bf 24}, R171 (2007)
  [gr-qc/0610018];
  %%CITATION = GR-QC/0610018;%%
%\cite{Percacci:2007sz}
%\bibitem{Percacci:2007sz} 
  R.~Percacci,
  %``Asymptotic Safety,''
  In Oriti, D. (ed.): ``Approaches to quantum gravity'' 111-128
  [arXiv:0709.3851 [hep-th]];
  %%CITATION = ARXIV:0709.3851;%%  
%\cite{Litim:2008tt}
%\bibitem{Litim:2008tt} 
  D.~F.~Litim,
  %``Fixed Points of Quantum Gravity and the Renormalisation Group,''
  arXiv:0810.3675 [hep-th];
  %%CITATION = ARXIV:0810.3675;%%  
%\cite{Litim:2011cp}
%\bibitem{Litim:2011cp} 
  D.~F.~Litim,
  %``Renormalisation group and the Planck scale,''
  Phil.\ Trans.\ Roy.\ Soc.\ Lond.\ A {\bf 369}, 2759 (2011)
  [arXiv:1102.4624 [hep-th]];
  %%CITATION = ARXIV:1102.4624;%%
%\cite{Percacci:2011fr}
%\bibitem{Percacci:2011fr} 
  R.~Percacci,
  %``A Short introduction to asymptotic safety,''
  arXiv:1110.6389 [hep-th];
  %%CITATION = ARXIV:1110.6389;%%
%\cite{Reuter:2012id}
%\bibitem{Reuter:2012id} 
  M.~Reuter and F.~Saueressig,
  %``Quantum Einstein Gravity,''
  New J.\ Phys.\  {\bf 14}, 055022 (2012)
  [arXiv:1202.2274 [hep-th]];
  %%CITATION = ARXIV:1202.2274;%%
%\cite{Reuter:2012xf}
%\bibitem{Reuter:2012xf} 
  M.~Reuter and F.~Saueressig,
  %``Asymptotic Safety, Fractals, and Cosmology,''
  arXiv:1205.5431 [hep-th];
  %%CITATION = ARXIV:1205.5431;%%  
%
%\bibitem{Nagy:2012ef} 
  S.~Nagy,
  %``Lectures on renormalization and asymptotic safety,''
 Annals Phys.\  {\bf 350}, 310 (2014)
  doi:10.1016/j.aop.2014.07.027
  [arXiv:1211.4151 [hep-th]];
    A.~Ashtekar, M.~Reuter and C.~Rovelli,
  %``From General Relativity to Quantum Gravity,''
  arXiv:1408.4336 [gr-qc];
  %%CITATION = ARXIV:1408.4336;%%
  %\cite{Bonanno:2017pkg}
%\bibitem{Bonanno:2017pkg} 
  A.~Bonanno and F.~Saueressig,
  %``Asymptotically safe cosmology - a status report,''
  arXiv:1702.04137 [hep-th].
  %%CITATION = ARXIV:1702.04137;%%
   
\bibitem{cosmology}
%\cite{Bonanno:2000ep}
%\bibitem{Bonanno:2000ep} 
  A.~Bonanno and M.~Reuter,
  %``Renormalization group improved black hole space-times,''
  Phys.\ Rev.\ D {\bf 62}, 043008 (2000)
  doi:10.1103/PhysRevD.62.043008
  [hep-th/0002196];
  %%CITATION = doi:10.1103/PhysRevD.62.043008;%%
%\cite{Bonanno:2006eu}
%\bibitem{Bonanno:2006eu} 
  A.~Bonanno and M.~Reuter,
  %``Spacetime structure of an evaporating black hole in quantum gravity,''
  Phys.\ Rev.\ D {\bf 73}, 083005 (2006)
  doi:10.1103/PhysRevD.73.083005
  [hep-th/0602159];
  %%CITATION = doi:10.1103/PhysRevD.73.083005;%%
  %\cite{Falls:2010he}
%\bibitem{Falls:2010he} 
  K.~Falls, D.~F.~Litim and A.~Raghuraman,
  %``Black Holes and Asymptotically Safe Gravity,''
  Int.\ J.\ Mod.\ Phys.\ A {\bf 27}, 1250019 (2012)
  doi:10.1142/S0217751X12500194
  [arXiv:1002.0260 [hep-th]];
  %%CITATION = doi:10.1142/S0217751X12500194;%%
%\cite{Bonanno:2010bt}
%\bibitem{Bonanno:2010bt} 
  A.~Bonanno, A.~Contillo and R.~Percacci,
  %``Inflationary solutions in asymptotically safe f(R) theories,''
  Class.\ Quant.\ Grav.\  {\bf 28}, 145026 (2011)
  doi:10.1088/0264-9381/28/14/145026
  [arXiv:1006.0192 [gr-qc]];
  %%CITATION = doi:10.1088/0264-9381/28/14/145026;%%
  %\cite{Falls:2012nd}
%\bibitem{Falls:2012nd} 
  K.~Falls and D.~F.~Litim,
  %``Black hole thermodynamics under the microscope,''
  Phys.\ Rev.\ D {\bf 89}, 084002 (2014)
  doi:10.1103/PhysRevD.89.084002
  [arXiv:1212.1821 [gr-qc]];
  %%CITATION = doi:10.1103/PhysRevD.89.084002;%%
%\cite{Bonanno:2012jy}
%\bibitem{Bonanno:2012jy} 
  A.~Bonanno,
  %``An effective action for asymptotically safe gravity,''
  Phys.\ Rev.\ D {\bf 85}, 081503 (2012)
  doi:10.1103/PhysRevD.85.081503
  [arXiv:1203.1962 [hep-th]];
  %%CITATION = doi:10.1103/PhysRevD.85.081503;%%
  %\cite{Koch:2013owa}
%\bibitem{Koch:2013owa} 
  B.~Koch and F.~Saueressig,
  %``Structural aspects of asymptotically safe black holes,''
  Class.\ Quant.\ Grav.\  {\bf 31}, 015006 (2014)
  doi:10.1088/0264-9381/31/1/015006
  [arXiv:1306.1546 [hep-th]];
  %%CITATION = doi:10.1088/0264-9381/31/1/015006;%%
  %\cite{Koch:2014cqa}
%\bibitem{Koch:2014cqa} 
  B.~Koch and F.~Saueressig,
  %``Black holes within Asymptotic Safety,''
  Int.\ J.\ Mod.\ Phys.\ A {\bf 29}, no. 8, 1430011 (2014)
  doi:10.1142/S0217751X14300117
  [arXiv:1401.4452 [hep-th]];
  %%CITATION = doi:10.1142/S0217751X14300117;%%
  %\cite{Koch:2014joa}
%\bibitem{Koch:2014joa} 
  B.~Koch, P.~Rioseco and C.~Contreras,
  %``Scale Setting for Self-consistent Backgrounds,''
  Phys.\ Rev.\ D {\bf 91}, no. 2, 025009 (2015)
  doi:10.1103/PhysRevD.91.025009
  [arXiv:1409.4443 [hep-th]];
  %%CITATION = doi:10.1103/PhysRevD.91.025009;%%
%\cite{Bonanno:2015fga}
%\bibitem{Bonanno:2015fga} 
  A.~Bonanno and A.~Platania,
  %``Asymptotically safe inflation from quadratic gravity,''
  Phys.\ Lett.\ B {\bf 750}, 638 (2015)
  doi:10.1016/j.physletb.2015.10.005
  [arXiv:1507.03375 [gr-qc]];
  %%CITATION = doi:10.1016/j.physletb.2015.10.005;%%
%\cite{Bonanno:2016dyv}
%\bibitem{Bonanno:2016dyv} 
  A.~Bonanno, B.~Koch and A.~Platania,
  %``Cosmic Censorship in Quantum Einstein Gravity,''
  arXiv:1610.05299 [gr-qc].
  %%CITATION = ARXIV:1610.05299;%%       
       
%\cite{Eichhorn:2016vvy}
\bibitem{Eichhorn:2016vvy} 
  A.~Eichhorn and S.~Lippoldt,
  %``Quantum gravity and Standard-Model-like fermions,''
  Phys.\ Lett.\ B {\bf 767}, 142 (2017)
  doi:10.1016/j.physletb.2017.01.064
  [arXiv:1611.05878 [gr-qc]].
  %%CITATION = doi:10.1016/j.physletb.2017.01.064;%%


  
  
      %\cite{Narain:2009fy}
\bibitem{Narain:2009fy} 
  G.~Narain and R.~Percacci,
  %``Renormalization Group Flow in Scalar-Tensor Theories. I,''
  Class.\ Quant.\ Grav.\  {\bf 27}, 075001 (2010)
  doi:10.1088/0264-9381/27/7/075001
  [arXiv:0911.0386 [hep-th]].
  %%CITATION = doi:10.1088/0264-9381/27/7/075001;%%
  
  %\cite{Narain:2009gb}
\bibitem{Narain:2009gb} 
  G.~Narain and C.~Rahmede,
  %``Renormalization Group Flow in Scalar-Tensor Theories. II,''
  Class.\ Quant.\ Grav.\  {\bf 27}, 075002 (2010)
  doi:10.1088/0264-9381/27/7/075002
  [arXiv:0911.0394 [hep-th]].
  %%CITATION = doi:10.1088/0264-9381/27/7/075002;%%
  
  %
\bibitem{Gies:2002af} 
  H.~Gies,
  %``Running coupling in Yang-Mills theory: A flow equation study,''
  Phys.\ Rev.\ D {\bf 66}, 025006 (2002)
  doi:10.1103/PhysRevD.66.025006
  [hep-th/0202207].
  %%CITATION = doi:10.1103/PhysRevD.66.025006;%%
  
  %\cite{Eichhorn:2010zc}
\bibitem{Eichhorn:2010zc} 
  A.~Eichhorn, H.~Gies and J.~M.~Pawlowski,
  %``Gluon condensation and scaling exponents for the propagators in Yang-Mills theory,''
  Phys.\ Rev.\ D {\bf 83}, 045014 (2011)
  Erratum: [Phys.\ Rev.\ D {\bf 83}, 069903 (2011)]
  doi:10.1103/PhysRevD.83.069903, 10.1103/PhysRevD.83.045014
  [arXiv:1010.2153 [hep-ph]].
  %%CITATION = doi:10.1103/PhysRevD.83.069903, 10.1103/PhysRevD.83.045014;%%
  
  %\cite{Shaposhnikov:2009pv}
\bibitem{Shaposhnikov:2009pv} 
  M.~Shaposhnikov and C.~Wetterich,
  %``Asymptotic safety of gravity and the Higgs boson mass,''
  Phys.\ Lett.\ B {\bf 683}, 196 (2010)
  doi:10.1016/j.physletb.2009.12.022
  [arXiv:0912.0208 [hep-th]].
  %%CITATION = doi:10.1016/j.physletb.2009.12.022;%%
  
    \bibitem{EFT}
  %\cite{Donoghue:1994dn}
%\bibitem{Donoghue:1994dn} 
  J.~F.~Donoghue,
  %``General relativity as an effective field theory: The leading quantum corrections,''
  Phys.\ Rev.\ D {\bf 50}, 3874 (1994)
  doi:10.1103/PhysRevD.50.3874
  [gr-qc/9405057];
  %%CITATION = doi:10.1103/PhysRevD.50.3874;%%
   J.~F.~Donoghue,
  %``Leading quantum correction to the Newtonian potential,''
  Phys.\ Rev.\ Lett.\  {\bf 72}, 2996 (1994)
  doi:10.1103/PhysRevLett.72.2996
  [gr-qc/9310024];
  %%CITATION = doi:10.1103/PhysRevLett.72.2996;%%
  N.~E.~J.~Bjerrum-Bohr, J.~F.~Donoghue and B.~R.~Holstein,
  %``Quantum gravitational corrections to the nonrelativistic scattering potential of two masses,''
  Phys.\ Rev.\ D {\bf 67}, 084033 (2003)
  Erratum: [Phys.\ Rev.\ D {\bf 71}, 069903 (2005)]
  doi:10.1103/PhysRevD.71.069903, 10.1103/PhysRevD.67.084033
  [hep-th/0211072];
  %%CITATION = doi:10.1103/PhysRevD.71.069903, 10.1103/PhysRevD.67.084033;%%
  %\cite{Bjerrum-Bohr:2016hpa}
%\bibitem{Bjerrum-Bohr:2016hpa} 
  N.~E.~J.~Bjerrum-Bohr, J.~F.~Donoghue, B.~R.~Holstein, L.~Plante and P.~Vanhove,
  %``Light-like Scattering in Quantum Gravity,''
  JHEP {\bf 1611}, 117 (2016)
  doi:10.1007/JHEP11(2016)117
  [arXiv:1609.07477 [hep-th]].
  %%CITATION = doi:10.1007/JHEP11(2016)117;%%
  
  \bibitem{EFTabelian}
  %\cite{Robinson:2005fj}
%\bibitem{Robinson:2005fj} 
  S.~P.~Robinson and F.~Wilczek,
  %``Gravitational correction to running of gauge couplings,''
  Phys.\ Rev.\ Lett.\  {\bf 96}, 231601 (2006)
  doi:10.1103/PhysRevLett.96.231601
  [hep-th/0509050];
  %%CITATION = doi:10.1103/PhysRevLett.96.231601;%%
  %\cite{Pietrykowski:2006xy}
%\bibitem{Pietrykowski:2006xy} 
  A.~R.~Pietrykowski,
  %``Gauge dependence of gravitational correction to running of gauge couplings,''
  Phys.\ Rev.\ Lett.\  {\bf 98}, 061801 (2007)
  doi:10.1103/PhysRevLett.98.061801
  [hep-th/0606208];
  %%CITATION = doi:10.1103/PhysRevLett.98.061801;%%
  %\cite{Ebert:2007gf}
%\bibitem{Ebert:2007gf} 
  D.~Ebert, J.~Plefka and A.~Rodigast,
  %``Absence of gravitational contributions to the running Yang-Mills coupling,''
  Phys.\ Lett.\ B {\bf 660}, 579 (2008)
  doi:10.1016/j.physletb.2008.01.037
  [arXiv:0710.1002 [hep-th]];
  %%CITATION = doi:10.1016/j.physletb.2008.01.037;%%
  %\cite{Toms:2007sk}
%\bibitem{Toms:2007sk} 
  D.~J.~Toms,
  %``Quantum gravity and charge renormalization,''
  Phys.\ Rev.\ D {\bf 76}, 045015 (2007)
  doi:10.1103/PhysRevD.76.045015
  [arXiv:0708.2990 [hep-th]];
  %%CITATION = doi:10.1103/PhysRevD.76.045015;%%
  %\cite{Toms:2008dq}
%\bibitem{Toms:2008dq} 
  D.~J.~Toms,
  %``Cosmological constant and quantum gravitational corrections to the running fine structure constant,''
  Phys.\ Rev.\ Lett.\  {\bf 101}, 131301 (2008)
  doi:10.1103/PhysRevLett.101.131301
  [arXiv:0809.3897 [hep-th]];
  %%CITATION = doi:10.1103/PhysRevLett.101.131301;%%
  %\cite{Toms:2009vd}
%\bibitem{Toms:2009vd} 
  D.~J.~Toms,
  %``Quantum gravity, gauge coupling constants, and the cosmological constant,''
  Phys.\ Rev.\ D {\bf 80}, 064040 (2009)
  doi:10.1103/PhysRevD.80.064040
  [arXiv:0908.3100 [hep-th]];
  %%CITATION = doi:10.1103/PhysRevD.80.064040;%%
  %\cite{Toms:2010vy}
%\bibitem{Toms:2010vy} 
  D.~J.~Toms,
  %``Quantum gravitational contributions to quantum electrodynamics,''
  Nature {\bf 468}, 56 (2010)
  doi:10.1038/nature09506
  [arXiv:1010.0793 [hep-th]];
  %%CITATION = doi:10.1038/nature09506;%%
  %\cite{Toms:2011zza}
%\bibitem{Toms:2011zza} 
  D.~J.~Toms,
  %``Quadratic divergences and quantum gravitational contributions to gauge coupling constants,''
  Phys.\ Rev.\ D {\bf 84}, 084016 (2011).
  doi:10.1103/PhysRevD.84.084016
  %%CITATION = doi:10.1103/PhysRevD.84.084016;%%
  
  
  \bibitem{AS4d}
  %\cite{Litim:2014uca}
%\bibitem{Litim:2014uca} 
  D.~F.~Litim and F.~Sannino,
  %``Asymptotic safety guaranteed,''
  JHEP {\bf 1412}, 178 (2014)
  doi:10.1007/JHEP12(2014)178
  [arXiv:1406.2337 [hep-th]];
  %%CITATION = doi:10.1007/JHEP12(2014)178;%%
  %\cite{Intriligator:2015xxa}
%\bibitem{Intriligator:2015xxa} 
  K.~Intriligator and F.~Sannino,
  %``Supersymmetric asymptotic safety is not guaranteed,''
  JHEP {\bf 1511}, 023 (2015)
  doi:10.1007/JHEP11(2015)023
  [arXiv:1508.07411 [hep-th]];
  %%CITATION = doi:10.1007/JHEP11(2015)023;%%
%\cite{Esbensen:2015cjw}
%\bibitem{Esbensen:2015cjw} 
  J.~K.~Esbensen, T.~A.~Ryttov and F.~Sannino,
  %``Quantum critical behavior of semisimple gauge theories,''
  Phys.\ Rev.\ D {\bf 93}, no. 4, 045009 (2016)
  doi:10.1103/PhysRevD.93.045009
  [arXiv:1512.04402 [hep-th]];
  %%CITATION = doi:10.1103/PhysRevD.93.045009;%%
  %\cite{Bond:2016dvk}
%\bibitem{Bond:2016dvk} 
  A.~D.~Bond and D.~F.~Litim,
  %``Theorems for Asymptotic Safety of Gauge Theories,''
  arXiv:1608.00519 [hep-th];
  %%CITATION = ARXIV:1608.00519;%%
  %\cite{Bajc:2016efj}
%\bibitem{Bajc:2016efj} 
  B.~Bajc and F.~Sannino,
  %``Asymptotically Safe Grand Unification,''
  JHEP {\bf 1612}, 141 (2016)
  doi:10.1007/JHEP12(2016)141
  [arXiv:1610.09681 [hep-th]];
  %%CITATION = doi:10.1007/JHEP12(2016)141;%%
  %\cite{Pelaggi:2017wzr}
%\bibitem{Pelaggi:2017wzr} 
  G.~M.~Pelaggi, F.~Sannino, A.~Strumia and E.~Vigiani,
  %``Naturalness of asymptotically safe Higgs,''
  arXiv:1701.01453 [hep-ph];
  %%CITATION = ARXIV:1701.01453;%%
  %\cite{Bond:2017wut}
%\bibitem{Bond:2017wut} 
  A.~D.~Bond, G.~Hiller, K.~Kowalska and D.~F.~Litim,
  %``Directions for model building from asymptotic safety,''
  arXiv:1702.01727 [hep-ph].
  %%CITATION = ARXIV:1702.01727;%%
  
  \bibitem{Christiansen2017}
  N.~Christiansen, A.~Eichhorn and A.~Held,
  in preparation.
  
  %
\bibitem{Daum:2009dn} 
  J.~E.~Daum, U.~Harst and M.~Reuter,
  %``Running Gauge Coupling in Asymptotically Safe Quantum Gravity,''
  JHEP {\bf 1001}, 084 (2010)
  doi:10.1007/JHEP01(2010)084
  [arXiv:0910.4938 [hep-th]].
  %%CITATION = doi:10.1007/JHEP01(2010)084;%%
  
  %\cite{Folkerts:2011jz}
\bibitem{Folkerts:2011jz} 
  S.~Folkerts, D.~F.~Litim and J.~M.~Pawlowski,
  %``Asymptotic freedom of Yang-Mills theory with gravity,''
  Phys.\ Lett.\ B {\bf 709}, 234 (2012)
  doi:10.1016/j.physletb.2012.02.002
  [arXiv:1101.5552 [hep-th]].
  %%CITATION = doi:10.1016/j.physletb.2012.02.002;%%

  \bibitem{ChristiansenLitimPawlowski2017}
  N.~Christiansen, D.~F.~Litim and J.~M.~Pawlowski,
  in preparation.
  
  %\cite{Peskin:1980ay}
\bibitem{Peskin:1980ay} 
  M.~E.~Peskin,
  %``Critical Point Behavior Of The Wilson Loop,''
  Phys.\ Lett.\  {\bf 94B}, 161 (1980).
  doi:10.1016/0370-2693(80)90848-5
  %%CITATION = doi:10.1016/0370-2693(80)90848-5;%%
  
  %\cite{Gies:2003ic}
\bibitem{Gies:2003ic} 
  H.~Gies,
  %``Renormalizability of gauge theories in extra dimensions,''
  Phys.\ Rev.\ D {\bf 68}, 085015 (2003)
  doi:10.1103/PhysRevD.68.085015
  [hep-th/0305208].
  %%CITATION = doi:10.1103/PhysRevD.68.085015;%%
  
\end{thebibliography}
\end{document}